\documentclass[aps,pre]{revtex4}  
\usepackage{epsf,moreverb,amsmath}
\usepackage{amsmath}
\usepackage{graphicx}

\begin{document}

\title{Introduction to  Q-tensor theory}
\author{N.J.~Mottram}
\email[author for correspondence:\ \ ]{nigel.mottram@strath.ac.uk}
\affiliation{Department of Mathematics and Statistics, University of Strathclyde, Livingstone Tower, Glasgow G1 1XH, U.K.}
\author{C.J.P. Newton}
\affiliation{Peartree Cottage, Little London, Longhope, Gloucestershire GL17 0PH, U.K.}
 
\begin{abstract}
This paper aims to provide an introduction to a basic form of the ${\bf Q}$-tensor approach to modelling liquid crystals, which has seen increased interest in recent years. The increase in interest in this type of modelling approach has been driven by investigations into the fundamental nature of defects and new applications of liquid crystals such as bistable displays and colloidal systems for which a description of defects and disorder is essential. The work in this paper is not new research, rather it is an introductory guide for anyone wishing to model a system using such a theory. A more complete mathematical description of this theory, including a description of flow effects, can be found in numerous sources but the books by Virga and Sonnet  and Virga are recommended. More information can be obtained from the plethora of papers using such approaches, although a general introduction for the novice is lacking. The first  few sections of this paper will detail the development of the ${\bf Q}$-tensor approach for nematic liquid crystalline systems and construct the free energy and governing equations for the mesoscopic dependent variables. A number of device surface treatments are considered and theoretical boundary conditions are specified for each instance. Finally, an example of a real device is demonstrated.
\end{abstract}

\pacs{}
\maketitle

\section{Background}
\label{background} In all matter, the physical state of the material can be described in terms of the translational and
rotational motion of the constituent molecules. In a crystalline solid the intermolecular forces hinder motion and force the molecules to lie, on average, in a regular array (i.e.~in a crystal lattice structure). As the substance is heated the molecules gain kinetic energy and large molecular vibrations break the crystal structure resulting in a fluid phase. In the liquid fluid phase the intermolecular forces are still strong over the average distance between molecules but cannot maintain the crystal structure. In the gas fluid phase the average distance between molecules is large, the intermolecular forces are weak and the unrestricted motion of the molecules cause the gas to expand and fill the container holding it.

\begin{figure}
\noindent \leavevmode \centerline{ 
  \includegraphics[width=8cm]{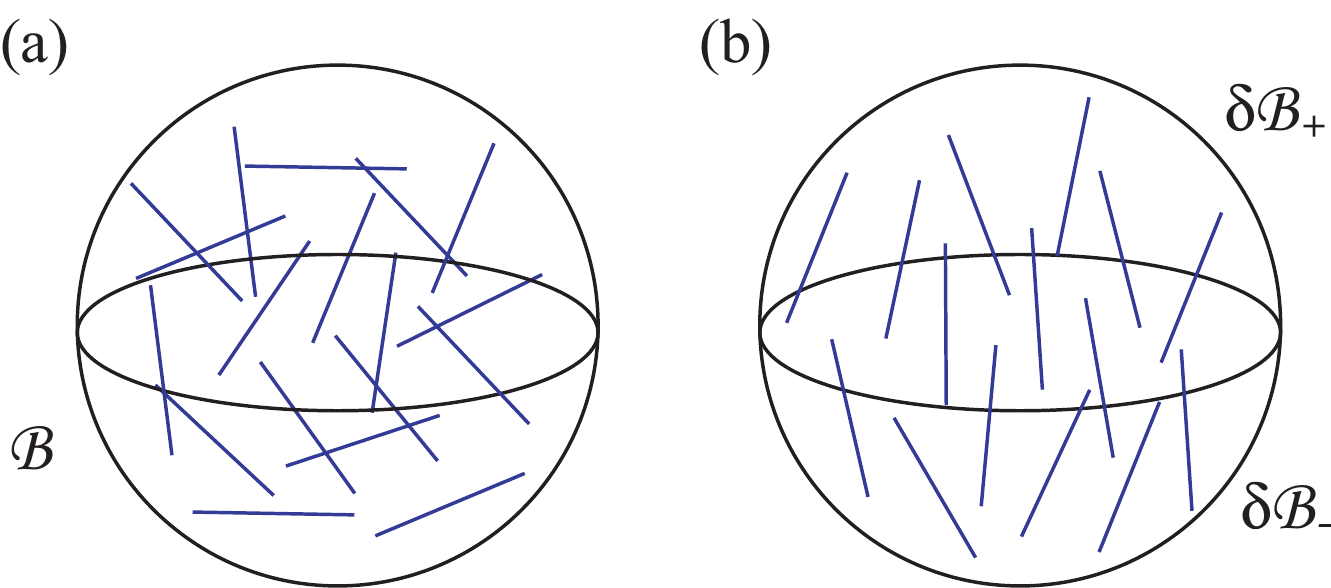}} \caption{A snap-shot of molecules within the
  ball ${\cal B}$ in (a) the isotropic liquid phase and (b) the nematic liquid crystal
  phase. Due to the head-tail symmetry of the molecules, the order
  parameter integration need only be performed over the  half sphere
  $\delta{\cal B}_+$ or $\delta{\cal B}_-$.} \label{phases}
\end{figure}

In certain materials, which typically consist of either rod-like or disc-like molecules, it is not just the translational motion, i.e.~the motion of positions of the centres of mass of the molecules, which determines the state of matter. We must also consider the orientation of the rod- or disc-like molecules. In a ``normal'' liquid, or more correctly, an isotropic liquid, the orientation of the molecule is random. If we pick a single molecule and consider how the orientation varies with time it will eventually take all possible orientations. Alternatively, if we look in a small region of the material (say a fixed ball, ${\cal
  B}$,  centred at the point ${\bf x}$, see Fig.~\ref{phases}(a)), then the molecules in the ball will be randomly oriented. Such a
system, in which a time average is equivalent to a space average, is said to be {\it ergodic}.

However, in a temperature region between the isotropic liquid and crystal states, some materials exhibit an amount of orientational order. That is, if we look in the small ball ${\cal B}$ the orientation is not random, rather there exists an average
orientation, see Fig.~\ref{phases}(b).  Materials which exhibit such orientational order are called {\it liquid crystals}.
\subsection{Measures of orientational order}
\label{order} 
\begin{figure}
\noindent \leavevmode \centerline{ 
  \includegraphics[width=15cm]{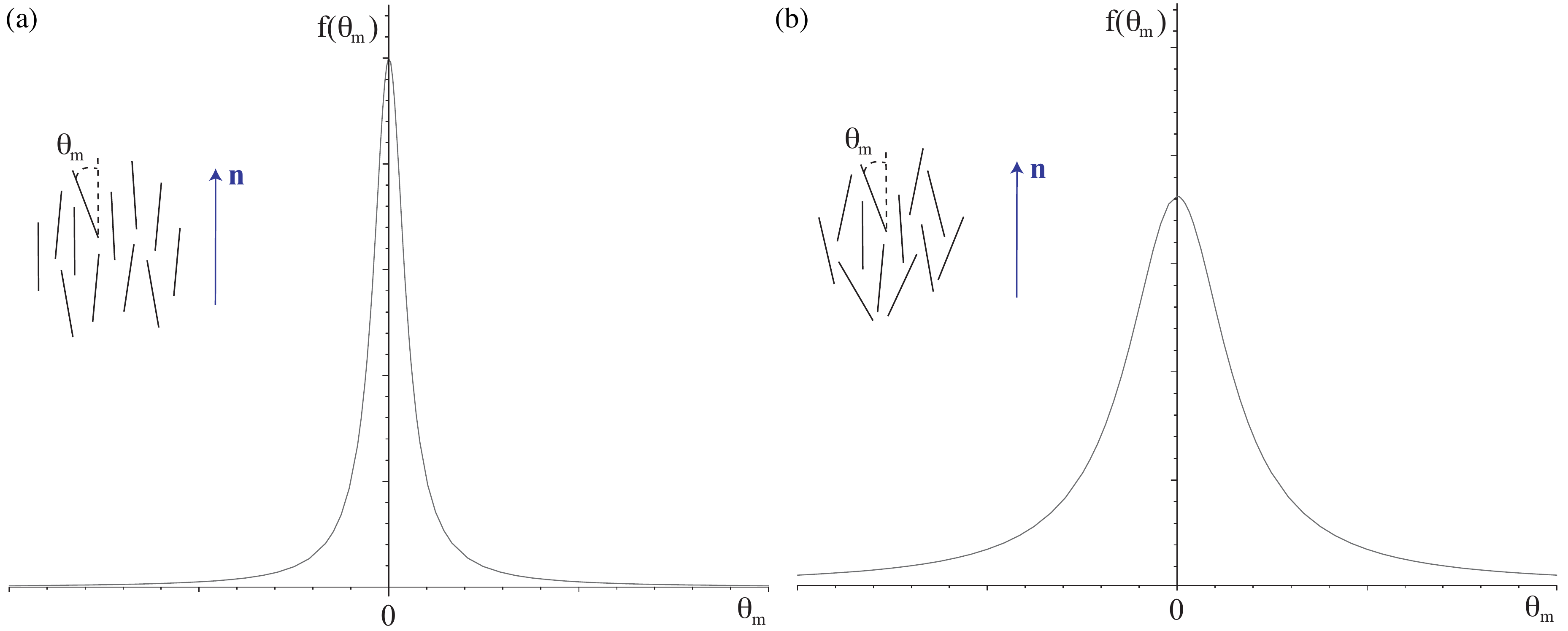}} \caption{(a) High orientational order: If   $\theta_m$ denotes the angle between a molecule and the director and  $f(\theta_m)$ is the probability of finding a molecule with  orientation $\theta_m$ in the small region ${\cal B}$ then for a  highly ordered system, $f(\theta_m)$ has a small standard  deviation. (b)Low orientational order: We can still  define the average orientation, i.e.~the mean of $f(\theta_m)$ but  in this case the molecules have more energy and the orientational  distribution is more spread out. The standard deviation of  $f(\theta_m)$ is larger.}
\label{dist1}
\end{figure}

For such a situation we can define a unit vector ${\bf n}$ called the {\it director} to be the average molecular orientation direction. For rod-like (or calamitic) molecules this will be the average long axis orientation and for disc-like (discotic) molecules it will be the average orientation of the disc normal. If we consider a different region in the material the director ${\bf n}$ may be in a different direction. This director may also vary with time and so it is a variable which is dependent on both space and time coordinates, ${\bf
  n}({\bf x},\,t)$. We can also measure the {\it amount} of orientational order about this average direction, in the small ball ${\cal B}$. If we consider all the molecules in ${\cal B}$ and construct a probability distribution of the molecular orientations then the mean of this
distribution is the director orientation but we can also compute, for instance, the standard deviation of the distribution. This
would measure how spread out the molecular orientation distribution is. However, rather than the standard deviation of
this distribution, the usual measure of this amount of order is the {\it scalar order parameter}, usually denoted by $S$. This is
a weighted average of the molecular orientation angles $\theta_m$ between the long molecular axes and the director
\begin{equation}
\label{Sdef1} S=\frac{1}{2}<3\cos^2\theta_m-1>,
\end{equation}
where $<>$ denotes the thermal or statistical average so that
\begin{equation}
\label{Sdef2} S=\frac{1}{2}\int_{{\cal
B}}(3\cos^2\theta_m-1)\,f(\theta_m)\,dV,
\end{equation}
and $f(\theta_m)$ is the statistical distribution function of the molecular angle $\theta_m$. Figure~\ref{dist1}  shows typical statistical distribution functions $f(\theta_m)$ for two cases, high orientational order and low orientational order. The function $f(\theta_m)$ will be even and periodic due to the head-tail symmetry of the molecules which
implies $f(\theta_m+\pi)=f(\theta_m)$.

In the ball ${\cal B}$ each molecular orientation can be described by the vector parallel to the molecular long axis or alternatively
a point on the surface of a half unit sphere, the point which the molecule ``points at'' on the sphere. The integration in
eq.~(\ref{Sdef2}) may therefore be performed over half the boundary of the sphere, $\delta{\cal B}^+$ say in Fig.~\ref{phases}(b).
\begin{figure}
\noindent \leavevmode \centerline{ 
\includegraphics[width=5cm]{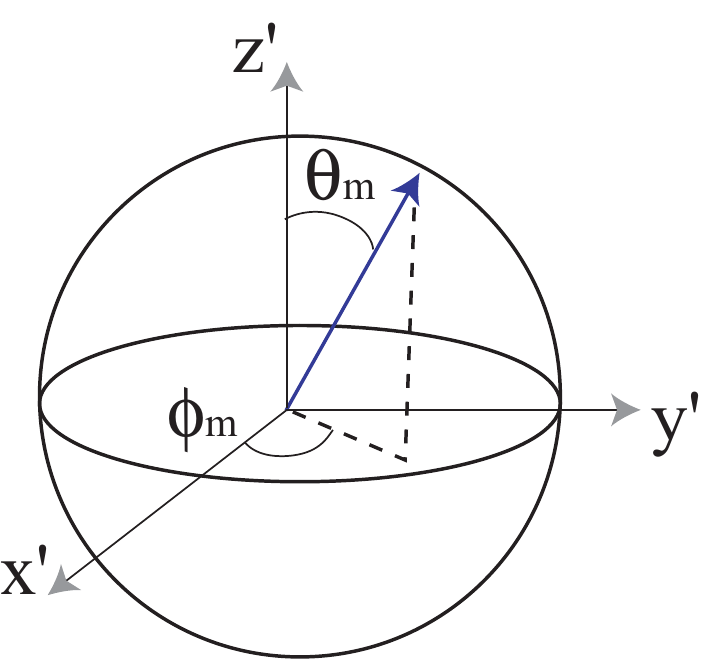}} \caption{A description of the orientation of a  single molecule in terms of the zenithal angle $\theta_m$ and  azimuthal angle $\phi_m$. The $x'$, $y'$ and $z'$ axes are locally defined and are not the laboratory frame of reference.}
\label{onemol}
\end{figure}
An even function of $\theta_m$ must be used in eq.~(\ref{Sdef1}) because, since $f(\theta_m)$ is an even function, any odd function of $\theta_m$ would lead to zero upon integration in eq.~(\ref{Sdef2}). To leading order this function is a good
approximate measure of the orientational order although a more accurate measure would include higher order polynomials of
$\cos\theta_m$. Equation~(\ref{Sdef1}) is in fact the second order Legendre polynomial, $P_2(\cos\theta_m)$, and it is usual to use the higher order even Legendre polynomials $P_{2n}(\cos\theta_m)$ when greater accuracy is required.

When the material is crystalline all molecules align exactly with the director and so $\theta_m=0$ for all molecules, which means that $<\cos^2\theta_m>=1$ and, from eq.~(\ref{Sdef1}), $S=1$. When all molecules lie in the plane perpendicular to the director, but randomly oriented in that plane, then the average still leads to the same director but $\theta_m=\pi/2$ for all molecules so that $<\cos^2\theta_m>=0$ and $S=-1/2$. In the isotropic liquid phase the molecules are
randomly oriented and so $f(\theta_m)$ is constant and equal to $1/(2\pi)$ (which can be derived from the property of probability distributions that $\int f(\theta_m)dV=1$). Therefore, performing the integration in eq.~(\ref{Sdef2}) in spherical coordinates where $\theta_m$ is the angle between the molecule and the director and $\phi_m$ is the azimuthal angle, i.e.~the angle between a fixed direction in the plane perpendicular to the director and the projection of the director onto that plane (see Fig.~\ref{onemol}), we obtain (using the substitution $x=\cos(\theta_m)$ on the second line),
\begin{eqnarray}
\int_{\delta{\cal
    B}_+}\cos^2\theta_m\,f(\theta_m)\,
    dA\,&&=\frac{1}{2\pi}\int_0^{2\pi}
    \int_0^{\pi/2}\cos^2\theta_m\,\sin\theta_md\theta_m\,d\phi_m\nonumber\\
&&=\int_0^{\pi/2}\cos^2\theta_m\,\sin\theta_md\theta_m
    =-\int_1^{0}
x^2\,dx=\frac{1}{3}
\end{eqnarray}
and so $<\cos^2\theta_m>=1/3$ and from eq.~(\ref{Sdef1}), $S=0$.

Although it is possible to achieve a molecular configuration for which $S$ is negative, i.e.~$-1/2<S<0$, it is more usual that in the equilibrium liquid crystal state the scalar order parameter is positive, $0<S<1$. As the temperature of the material changes the scalar order parameter will change from $S=0$ in the isotropic state, at high temperature, to $S=1$ in the crystalline state, at low temperature. A typical scalar order parameter in the middle of the phase region, for this type of liquid crystal, might be $S=0.6$.

\subsection{Biaxial nematics}
\label{biaxial} The fundamental principle of any biaxial system is that there is no axis of complete rotational symmetry (i.e no axis about which a rotation of {\it any} angle leaves the system unchanged), unlike a uniaxial system which has an  axes of rotational symmetry (such as the director ${\bf n}$ in uniaxial liquid crystals). For instance, some solid materials (i.e.~calcite) are uniaxial, and therefore birefringent, so that light travelling along the optical axis sees a different refractve index than light travelling perpendicular to the optic axis. However, other materials are biaxial (i.e.~olivine), and have trirefringence, where there are three different refractive indices, for light travelling along three orthogonal directions.

For such biaxial materials there can be defined a set of perpendicular axes (only two need to be defined as the third is then specified as perpendicular to the other two) for each of which there is a reflection symmetry. In biaxial liquid crystals the two axes ${\bf n}$ and ${\bf m}$ are therefore defined and the symmetries are the reflections ${\bf n}\rightarrow -{\bf n}$ and ${\bf m}\rightarrow -{\bf m}$.

The definition of three axes of reflective symmetry in a material is a relatively straightforward concept.  Less straightforward is to understand the molecular arrangement corresponding to a biaxial systems. There are two possibilities for the shape of the molecules: either the molecules are uniaxial or they are biaxial. A uniaxial molecule can be thought of as shaped like a cylinder or rod (uniaxial because there is a rotation symmetry about the axis of the molecule), a biaxial molecule might be shaped like a plank of wood. There is no rotation axis along the long axis of the plank (rotating the plank changes its configuration) but there are three axes of reflective symmetry (the long, intermediate and short axis of the plank).

We can now have a uniaxial or biaxial arrangement of uniaxial molecules and a uniaxial or biaxial arrangement of biaxial molecules. The easiest arrangements to imagine are the uniaxial arrangement of uniaxial molecules - a group of cylindrical molecules oriented, on average, along a single direction where rotation about this direction does not alter, at least statistically speaking, the arrangement - and also the biaxial arrangement of biaxial molecules, a group of plank-shaped molecules where the long axes of the planks align in one direction  and the short axes also align, in a second direction, so creating a biaxial ensemble arrangement.

A uniaxial arrangement of biaxial molecules is also relatively easy to imagine. In this arrangement the molecular long axes orient as before but  the short axes of the plank are oriented randomly so that, if we view from the side as in Fig.~\ref{phases}, at any moment in time some of the planks are facing out of the page and some are turned end on. When you rotate the system the individual molecules look different but on average the same proportion of molecules are face on and end on (and all orientations in between).

A more difficult situation to imagine is a biaxial arrangement of uniaxial molecules. Such an alignment is described below and is the most relevant to the liquid crystal structure at the centre of defects.

\begin{figure}
\noindent \leavevmode \centerline{
\includegraphics[width=9cm]{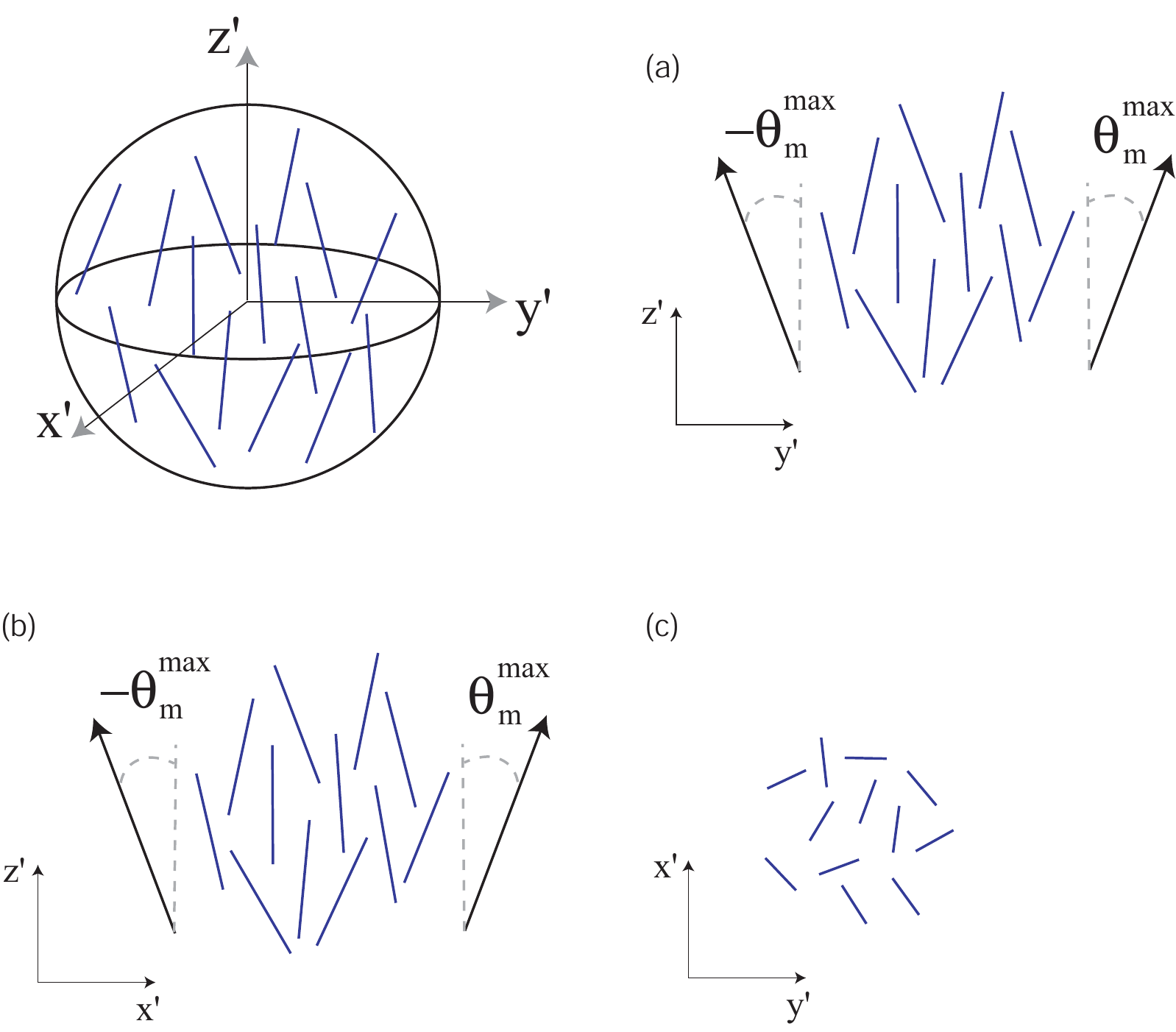}} \caption{A uniaxial  distribution of molecules.  Viewed along the $x'$ or $y'$-axes the molecular distribution is similar and the calculated scalar order parameter is identical. Viewed along the director, the $z'$-axis, the molecules are viewed virtually end-on and the distribution looks isotropic.} \label{uniax}
\end{figure}

\begin{figure}
\noindent \leavevmode \centerline{
\includegraphics[width=9cm]{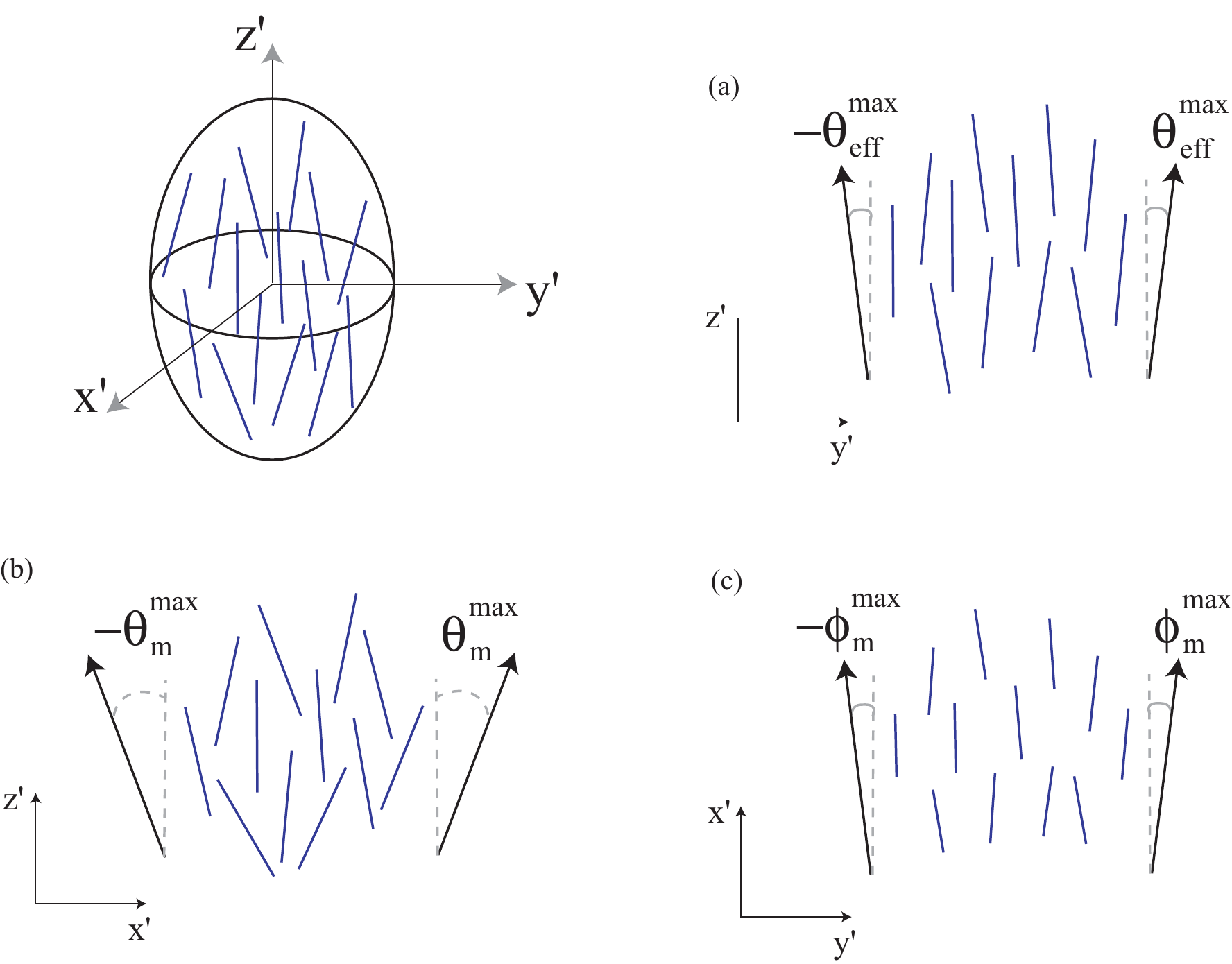}} \caption{A biaxial distribution of molecules.
Viewed along the three local axes the scalar order parameter, i.e.~how spread out the molecules are, is different in each case}
\label{biax}
\end{figure}

Imagine a group of uniaxial molecules (the cylindrical or rod-like molecules) which are in a uniaxial arrangement (Fig.~\ref{uniax}) if you look down the director (the $z'$ axis) the molecules, which are now viewed almost end-on, look random, it is only the side view (along the $x'$ or  $y'$ axis) which indicates the director ${\bf n}$. Now restrict this group of molecules by ``squashing'' the distribution within the ball (see Fig.~\ref{biax}) between two imaginary plane surfaces parallel to the $x'z'$ plane, and look along the $z'$ axis again. There will be a restriction of the motion of the molecules in Fig.~\ref{uniax}(c) to an arrangement similar to Fig.~\ref{biax}(c).

In more mathematical terminology: in the ball ${\cal B}$, if the distribution of molecules is such that the zenithal angles $\theta_m$ vary between $-\theta_m^{max}$ and $\theta_m^{max}$ and the azimuthal angles vary between $-\phi_m^{max}$ and $\phi_m^{max}$ then the three views of the group of molecules along the $x'$, $y'$ and $z'$ axes are shown in Fig.~\ref{biax}(a), (b) and (c) respectively. If we look at the molecules along the $y'$-axis (Fig.~\ref{biax}(b)) the extent of the zenithal angle is $-\theta_m^{max}<\theta_m<\theta_m^{max}$ and we can define a scalar order parameter $S_1$. If we look at the molecules along the $z'$-axis (Fig.~\ref{biax}(c)) the extent of the azimuthal angle is $-\phi_m^{max}<\phi_m<\phi_m^{max}$ and we can define a scalar order parameter $S_2$ say. If we look along the $x'$-axis (Fig.~\ref{biax}(a)), the restriction to the azimuthal angle means we see an effective zenithal angle range which is smaller than in Fig.~\ref{biax}(a). Simple geometry gives us that the {\it effective} range of zenithal angles is now
\begin{equation}
-\theta_{eff}^{max}=-\tan^{-1}\left(\sin\phi_m^{max}\tan\theta_m^{max}
\right)<\theta_m<\tan^{-1}\left(\sin\phi_m^{max}\tan\theta_m^{max}\right)
=\theta_{eff}^{max}
\end{equation}
Therefore, with this view, we would calculate a different scalar order parameter, $S_3$ say which is larger than $S_1$ (in fact
$S_3$ is related to $S_1$ and $S_2$). 
It is now necessary to define three directors and corresponding scalar order parameters. If we take two perpendicular directors ${\bf n}$ and ${\bf m}$, in our situation the $z'$ and $y'$ directions, then we can define the two scalar order parameters $S_1$ and $S_2$ associated with order about the two directors respectively (the third director being ${\bf n}\times{\bf m}$). If the order parameter $S_2$ is non-zero and not equal to $S_1$ the liquid crystal is said to be {\it biaxial}, i.e.~there are two axes of symmetry. If, however, $S_2$=0 then in comparison to Fig.~\ref{biax}(c) the range of azimuthal angles is now $-\pi/2<\phi_m<\pi/2$ and a view along the $z'$-axis shows randomly oriented molecules (see Fig.~\ref{uniax}). The view from the $x'$-axis will show a regular nematic distribution of molecules and the director and a non-zero order parameter $S_1$ can be defined. Since the system of molecules is now rotationally symmetric about the $z'$-axis, the view along the $y'$ axis will be the same as that along the $x'$-axis and thus the order parameter will again be $S_1$.

From these examples we see that the uniaxial nematic state exists when the order parameter with respect to one direction is zero, or equivalently the order parameters with respect to the perpendicular directions are the same. The biaxial nematic state
exists when the scalar order parameters in all three perpendicular directions are different.

It should be mentioned that a stable biaxial thermotropic nematic liquid crystal has only recently been reported (after over 50 years of searching). Many attempts to synthesise such a material, mainly using flat, board-like molecules, have failed and it seems that bent-core molecules have provided the solution \cite{biax1,biax2,Luckhurst:2012}. The biaxial state is certainly important since certain external forces such as surfaces in contact with the liquid crystal, electric fields and some topological constraints can induce a biaxial state in regions of a uniaxial nematic liquid crystal sample.

In this paper we will only consider an ensemble of rod-like (i.e. uniaxial) molecules which, as mentioned above, may form a uniaxial or biaxial configuration. Consideration of an ensemble of plank-like or bent core (i.e. biaxial) molecules is more complicated and involves a second order tensor ${\bf B}$ to describe the ordering of the short molecular axes. Such a theory is detailed in Sonnet and Virga \cite{SonnetVirga}.  

In summary, the most general nematic state is the biaxial state which can be described by two vector-valued variables and two
scalar-valued variables, the directors ${\bf n}({\bf x},\,t)$ and ${\bf m}({\bf x},\,t)$ and the scalar order parameters $S_1({\bf x},\,t)$ and $S_2({\bf x},\,t)$, all of which may depend on the spatial coordinates ${\bf x}$ and time.

We assume, without loss of generality, that the directors are of unit length, $|{\bf n}|=1$ and $|{\bf m}|=1$.
\begin{figure}
\noindent \leavevmode \centerline{ 
\includegraphics[width=4cm]{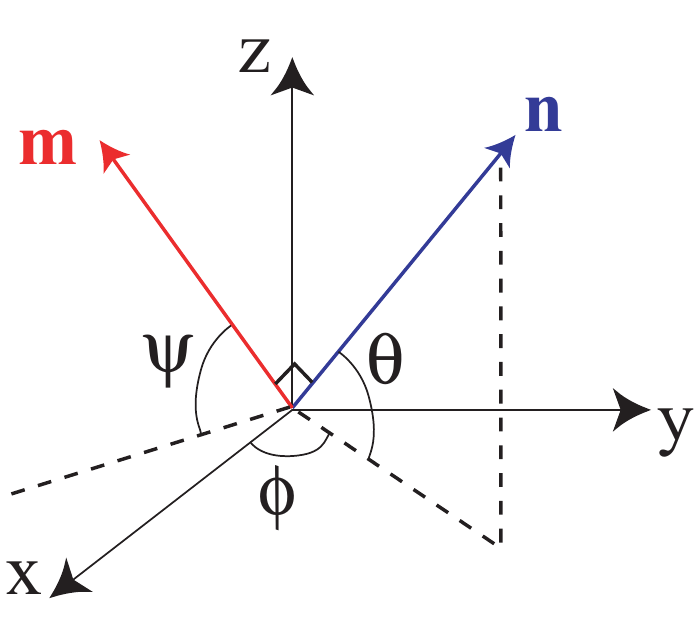}} \caption{The directors ${\bf n}$ and ${\bf m}$ in terms of the Euler angles $\theta$, $\phi$ and $\psi$. The axes $x$, $y$ and $z$ form the laboratory frame of reference.}
\label{director}
\end{figure}
Just as we have described each molecule in Fig.~\ref{onemol}, each of these directors can be represented in terms of the standard Euler angles, see Fig.~\ref{director}. Since the director ${\bf n}$ is of unit length it can be written as,
\begin{equation}
{\bf n}=\left(\cos\theta\cos\phi,\ \cos\theta\sin\phi,\
\sin\theta\right).
\end{equation}
It is important not to confuse the director angles $\theta$ and $\phi$ with the molecular angles in Fig.~\ref{onemol}. The molecular angles $\theta_m$ and $\phi_m$ are angles relative to a local frame of reference, i.e.~the directors, whereas the angles $\theta$ and $\phi$ are defined relative to the laboratory frame of reference, i.e a set of axes fixed in space, independent of the orientation of the material.

The director ${\bf m}$ is perpendicular to ${\bf n}$ and therefore has one remaining degree of freedom, the angle $\psi$,
\begin{equation}
{\bf m}=\left(\sin\phi\cos\psi-\cos\phi\sin\psi\sin\theta,\
-\sin\phi\sin\psi\sin\theta-\cos\phi\cos\psi,\
\sin\psi\cos\theta\right).
\end{equation}
The angle $\psi$ is the angle from ${\bf m}$ to the direction $(\sin\phi,\ -\cos\phi,\ 0)$ in the $xy$-plane which is also perpendicular to ${\bf n}$.

A theory can now be constructed using the {\it five} dependent variables,
\begin{equation}
\label{vars} \theta({\bf x},\,t),\ \phi({\bf x},\,t),\ \psi({\bf
x},\,t),\ S_1({\bf x},\,t),\ S_2({\bf x},\,t).
\end{equation}
However, there may be problems with a theory such as this, based on Euler angles. When the zenithal angle $\theta$ equals $\pi/2$ the azimuthal angle $\phi$ is undefined and, as with all such angle variables, there may also be a problem with multi-valuedness since $\phi=0$ is equivalent to $\phi=2\pi$. Care must be taken when solving governing differential equations for $\theta$, $\phi$ and $\psi$.

\subsection{The tensor order parameter ${\bf Q}$} 
\label{Qtensor} We now define an alternative approach which removes the problems of the angle representation described above. Instead of defining the nematic state in terms of the five separate variables in eq.~(\ref{vars}) we construct a 3x3 matrix which contains all the information about the nematic state, i.e.~the information contained in the five variables. The problems of solving the Euler angle governing equations will be removed with this approach (for reference see for example \cite{Virga, SonnetVirga}).

Consider the 3x3 matrix,
\begin{equation}
\label{m}
{\bf M}=S_1\left({\bf n}\otimes{\bf n}\right)+S_2\left({\bf m}\otimes{\bf m}\right),
\end{equation}
where, for any vector ${\bf h}$ the $ij^{\rm th}$ element of the
product  ${\bf h}\otimes{\bf h}$ is $h_i\,h_j$, the $i^{\rm th}$
element of ${\bf h}$ multiplied by the $j^{\rm th}$ element of
${\bf h}$. This matrix will be symmetric, since
$n_i\,n_j=n_j\,n_i$ and $m_i\,m_j=m_j\,m_i$, and the trace of
${\bf M}$ will be $S_1(n_1n_1+n_2n_2+n_3n_3)+S_2(m_1m_1+m_2m_2+m_3m_3)=S_1+S_2$ since $|{\bf n}|=1$ and $|{\bf m}|=1$

Since the trace of ${\bf M}$ is fixed and the matrix is symmetric, ${\bf M}$ will have five independent elements,
\begin{equation}
{\bf M}=\left(
\begin{array}{ccc}
m_1&m_2&m_3\\
m_2&m_4&m_5\\
m_3&m_5&(S_1+S_2)-m_1-m_4
\end{array}
\right)
\end{equation}
The tensor ${\bf M}$ contains the same information  as the five separate variables in eq.~(\ref{vars}). However, if we construct a theory using ${\bf M}$ there will be no problems with the degeneracies of Euler angles.

As will be indicated later, it is actually more useful to use the tensor,
\begin{equation}
\label{q} {\bf Q}=S_1\left({\bf n}\otimes{\bf
n}\right)+S_2\left({\bf m}\otimes{\bf
m}\right)-\frac{1}{3}(S_1+S_2){\bf I},
\end{equation}
where ${\bf I}$ is the  identity matrix so that the trace of ${\bf Q}$ is zero. The {\it tensor order parameter} ${\bf Q}$ is therefore a symmetric traceless matrix and can be written as,
\begin{equation}
\label{qmat}
{\bf Q}=\left(
\begin{array}{ccc}
q_1&q_2&q_3\\
q_2&q_4&q_5\\
q_3&q_5&-q_1-q_4
\end{array}
\right),
\end{equation}
and from the definitions of ${\bf n}$  and ${\bf m}$ and eqs~(\ref{q}) and (\ref{qmat}) we see that the elements $q_i$ can be written in terms of the variables $\theta$, $\phi$, $\psi$, $S_1$ and $S_2$ thus,
\begin{eqnarray}
q_1&=&S_{{1}}\cos^2\theta\cos^2\phi+S_{{2}}\left (\sin\phi\cos\psi-\cos\phi\sin\psi\sin\theta\right)^{2}-\frac{1}{3}(S_{{1}}+S_{{2}}),\label{qs1}\\
q_2&=&S_{{1}}\cos^2\theta\sin\phi\cos\phi\nonumber\\
&&-S_{{2}}\left(\cos\phi\cos\psi+\sin\phi\sin\psi\sin\theta\right)\left(\sin\phi\cos\psi-\cos\phi\sin\psi\sin\theta\right),\label{qs2}\\
q_3&=&S_{{1}}\sin\theta\cos\theta\cos\phi+S_{{2}}\sin\psi\cos\theta\left
(\sin\phi\cos\psi-\cos\phi\sin\psi\sin\theta\right),\label{qs3}
\\
q_4&=&S_{{1}}\cos^2\theta\sin^2\phi+S_{{2}}\left(\cos\phi\cos\psi+\sin\phi\sin\psi\sin\theta\right)^{2}-\frac{1}{3}(S_{{1}}+S_{{2}}),\label{qs4}\\
q_5&=&S_{{1}}\cos\theta\sin\theta\sin\phi-S_{{2}}\sin\psi\cos\theta\left(\cos\phi\cos\psi+\sin\phi\sin\psi\sin\theta\right).\label{qs5}
\end{eqnarray}

From the previous section, the description of biaxiality tells us that, if the matrix ${\bf Q}$ was diagonalised (so that the eigenvectors are along the $x'$, $y'$ and $z'$ directions) then for a uniaxial state two of the eigenvalues would be the same and for a biaxial state all three eigenvalues would be different.

The eigenvalues of the matrix ${\bf Q}$, described by
eqs~(\ref{qs1})-(\ref{qs5}), are,
\begin{eqnarray}
\lambda_1&=&(2S_1-S_2)/3,\\
\lambda_2&=&-(S_1+S_2)/3,\\
\lambda_3&=&(2S_2-S_1)/3.
\end{eqnarray}
Uniaxial states exist when two of these eigenvalues are the same i.e.~when $\lambda_1=\lambda_2$ so that $S_1=0$ or when $\lambda_2=\lambda_3$ so $S_2=0$ or when $\lambda_1=\lambda_3$ so $S_1=S_2$. When all the eigenvalues are the same we have an isotropic system and we have $S_1=0$ and $S_2=0$ so that ${\bf Q}={\bf 0}$. We shall see later that the use of ${\bf Q}$ rather than ${\bf M}$ as our dependent tensor variable was in fact dictated by the property that the isotropic state is described by ${\bf Q}={\bf 0}$.

If we, for simplicity, define our laboratory axes to be such that the directors ${\bf n}$ and ${\bf m}$ are in the directions of the $x$ and $y$ axes then $\theta=0$, $\phi=0$ and $\psi=\pi$ so that
\begin{eqnarray}
q_1&=&\frac{1}{3}(2S_{{1}}-S_{{2}}),\label{qs1a}\\
q_2&=&0,\label{qs2a}\\
q_3&=&0,\label{qs3a}
\\
q_4&=&\frac{1}{3}(2S_{{2}}-S_{{1}}),\label{qs4a}\\
q_5&=&0,\label{qs5a}
\end{eqnarray}
and therefore the three possible uniaxial states correspond to: $q_1=-q_1-q_4$ so that $q_1=-q_4/2=-S_{2}/3$; $-q_1-q_4=q_4$ so that $q_4=-q_1/2=-S_{1}/3$; or $q_1=q_4=S_1/3=S_2/3$.


There will be no problems when solving the governing equations for the five dependent variables $q_1,\ q_2,\ q_3,\ q_4,\ q_5$ because the Euler angles only appear in $\sin$ and $\cos$ functions and the problem with the multivaluedness of the angles is removed.

\section{Phenomenological theory}
\label{theory} In order to form a theoretical framework to consider such liquid crystals it is necessary to construct the total free energy $\cal F$ of the liquid crystal sample, which may include terms such as: the elastic energy of any distortion to the structure of the material; thermotropic energy which dictates the preferred phase of the material; electric and/or magnetic energy from an externally applied electric or magnetic field and, in polar materials, the internal self-interaction energy of the polar molecules; surface energy terms representing the interaction energy between the bounding surface and the liquid crystal molecules at the surface. The total energy is therefore,
\begin{eqnarray}
\label{totenergyQ} {\cal F}&=&{\cal F}_{distortion}+{\cal
F}_{thermotropic}+{\cal F}_{electromagnetic}+{\cal
F}_{surface}\nonumber\\
&&=\int_V{\left(F_{d}+F_{t}+F_{e}\right)\,{\rm
d}v}+\int_S{\left(F_{s}\right)\,{\rm d}s},
\end{eqnarray}
where the energy densities, $F_{d},\,F_t$ etc., depend on the dependent
variables, the tensor order parameter elements.

In static situations minimisation of this energy, using the calculus of variations, leads to sets of differential equations in the bulk of the material and at the surface, for each of the dependent variables. The solution of these bulk equations subject to the surface boundary conditions gives the equilibrium configuration of the dependent variables through the sample.

The free energy density is assumed to depend on the tensor order parameter ${\bf Q}$ and all first order differentials of ${\bf Q}$. It is also assumed that distortions of ${\bf Q}$ are small and therefore higher order differentials and high powers of first order differentials will be negligible. The bulk free energy density, $F_b=F_d+F_t+F_e$, will be an integral of a function dependent on the elements of ${\bf Q}$ and all derivatives of the elements whereas the surface free energy density is assumed to be a function of the elements of ${\bf Q}$ only,
\begin{equation}
\label{qen} {\cal F}=\int_V{F_b\left(q_i, \nabla\,q_i\right)\,{\rm
d}v}+\int_S{F_{s}\left(q_i\right)\,{\rm d}s}.
\end{equation}

{\bf Static equations}\\
The governing differential equations for $q_i({\bf x})$ are the Euler-Lagrange equations, the solution of which minimises the free energy. The general Euler-Lagrange equations for the free energy eq.~(\ref{qen}) are the five equations
\begin{equation} \label{qeq}  \sum_{j=1}^3\frac{\partial}{\partial x_j}\left(\frac{\partial
F_b}{\partial q_{i,j}}\right)=\frac{\partial F_b}{\partial q_i},
\end{equation}
for $i=1 \ldots 5$ and where $q_{i,j}=\partial q_i/\partial x_j$. These five equations should be solved together with suitable boundary conditions.

Equations~(\ref{qeq}) may be written as
\begin{equation} \label{qeqfem}
\nabla.{\pmb \Gamma^i}=f^i,
\end{equation}
where
\begin{eqnarray} \label{qeqfemg}
\Gamma^i_j&=& \frac{\partial
F_b}{\partial q_{i,j}},\\
f^i&=&\frac{\partial F_b}{\partial q_i}.
\end{eqnarray}

{\bf Dynamic equations}\\
When the dynamic evolution of the ${\bf Q}$ tensor is required and the coupling between the director rotations and fluid flow is thought to be small (an assumption that is not always true, but see \cite{SonnetVirga} for a full dynamical theory), a dissipation principle can be used to show that the
governing equations will be,
\begin{equation} \label{qeqfemdyn}
\gamma\frac{\partial {\cal{D}}}{\partial \dot{q}_i}=\nabla.{\pmb
\Gamma^i}-f^i,
\end{equation}
where ${\cal D}$ is the dissipation function
\begin{equation} \label{qeqfemdyn1}
{\cal{D}}={\rm tr}\,\left(\left(\frac{\partial {\bf Q}}{\partial
t}\right)^2\right),
\end{equation}
$\dot{q}_i=\partial q/\partial x_i$ and the viscosity $\gamma$ is
related to the standard nematic viscosity $\gamma_1$ (which is
equal to $\alpha_3-\alpha_2$ in Leslie viscosities) by,
\begin{equation} \label{qeqfemdyn2}
\gamma=\frac{\gamma_1}{S_{exp}},
\end{equation}
where $S_{exp}$ is the uniaxial order parameter of the liquid
crystal when the experimental measurement of $\gamma_1$ was taken.

{\bf Boundary conditions}\\
Two types of boundary conditions will be considered; strong
(infinite) or weak anchoring.

{{\bf (i)} Strong}: For strong, also termed infinite, anchoring we use Dirichlet
conditions. That is, the order tensor will be fixed at a specified
value at the domain boundary that has been dictated by some
substrate alignment technique. In this case the boundary condition
is
\begin{equation}
\label{stronganch} {\bf Q}={\bf Q}_s,
\end{equation}
where ${\bf Q}_s$ is the prescribed order tensor at the boundary. In this case there is no surface energy and $F_s=0$ in eq.~\ref{totenergyQ}.

{{\bf (ii)} Weak}: For a weak anchoring effect there exists a surface energy, $F_s$, which is added to the free energy and must be minimised at the same time as the bulk free energy. This minimisation leads to the condition that, on the boundary, the liquid crystal variables $q_i$ satisfy
\begin{equation}
\label{qeqsurf2} \sum_{j=1,2,3}\nu_j\left(\frac{\partial
F_b}{\partial q_{i,j}}\right)=\frac{\partial F_s}{\partial q_i}
\end{equation}
or equivalently
\begin{equation} \label{qeqfembc}
{\pmb \nu}.{\pmb \Gamma}^i=G^i
\end{equation}
where ${\pmb\Gamma}^i$ is as in eq.~(\ref{qeqfemg}),
$G^i=-\partial
  F_s/\partial q_i$ and where ${\pmb \nu }$ is the normal to the substrate. Usually the surface is planar or circular so that the surface normal is relatively simple, for example ${\pmb\nu}=(0,\,0,\,1)$ if the surface is a plane parallel to the $xy$-plane. However, in general the substrate normal is position dependent $\pmb{\nu}({\bf x})$.

In order to complete this theory it is necessary to specify each of the components of the free
energy. The following sections describe the thermotropic, elastic, electrostatic and surface energies.

\subsection{Landau-de Gennes thermotropic energy}
\label{LdeG} The thermotropic energy, $F_t$, is a potential function which dictates which state the liquid crystal would
prefer to be in, i.e.~a uniaxial state, a biaxial state or the isotropic state. At high temperature this potential should have a
minimum energy in the isotropic state, i.e.~${\bf Q}={\bf 0}$ whereas at low temperatures there should be minima at three
uniaxial states, i.e.~the states where any two of the eigenvalues of ${\bf Q}$ are equal. For rod-like molecules a bulk biaxial minimum energy state is not possible\cite{SonnetVirga}. The simplest form of such a function is a truncated Taylor expansion about ${\bf Q}={\bf 0}$
\cite{Schophol},
\begin{equation}
\label{thermo} F_t=a\,{\rm tr}\left({\bf
Q}^2\right)+\frac{2b}{3}\,{\rm tr}\left({\bf
Q}^3\right)+\frac{c}{2}\,\left({\rm tr}\left({\bf
Q}^2\right)\right)^2
\end{equation}
which is a quartic function of the $q_i$s. This energy is sometimes
written as
\begin{equation}
\label{thermo1} F_t=a\,{\rm tr}\left({\bf
Q}^2\right)+\frac{2b}{3}\,{\rm tr}\left({\bf Q}^3\right)+c\,{\rm
tr}\left({\bf Q}^4\right).
\end{equation}
Equations~(\ref{thermo}) and~(\ref{thermo1}) are equivalent only
if the factor of two in the $c$ term is included. This form of energy can also be thought of as an expansion in terms of the two invariants of the order tensor, ${\rm tr}\left({\bf Q}^2\right)$ and ${\rm det}({\bf Q})$.

The coefficients $a$, $b$ and $c$ will in general be temperature dependent although it is usual to approximate this dependency by assuming that $b$ and $c$ are independent of temperature whilst $a=\alpha(T-T^*)=\alpha\Delta T$ where $\alpha>0$ and $T^*$ is the fixed temperature at which the isotropic state becomes unstable.

As stated in Section~\ref{Qtensor}, the eigenvalues of ${\bf Q}$ are $\lambda_1=(2S_1-S_2)/3,\ \lambda_2=-(S_1+S_2)/3,\ \lambda_3=(2S_2-S_1)/3$ and since the trace of the n$^{\rm th}$ power of ${\bf Q}$ is the sum of the n$^{\rm th}$ powers of these eigenvalues, thus the thermotropic energy is simply a function of $S_1$ and $S_2$,
\begin{eqnarray}
\label{thermo2} F_t&=&a\sum_{i=1}^3{\lambda_i^2}+
\frac{2b}{3}\sum_{i=1}^3{\lambda_i^3}+
\frac{c}{2}\left(\sum_{i=1}^3{\lambda_i^2}\right)^2\\
&=&\frac{2a}{3}\left(S_1^2-S_1S_2+S_2^2\right)+
\frac{2b}{27}\left(2S_1^3-3S_1^2S_2-3S_1S_2^2+2S_2^3\right)\\
&&\hspace{0.5cm}+\frac{2c}{9}\left(
S_1^4-2S_1^3S_2+3S_1^2S_2^2-2S_1S_2^3+S_2^4 \right)\nonumber
\end{eqnarray}

For a uniaxial state such as $\lambda_2=\lambda_3$, or equivalently $S_2=0$, the thermotropic energy becomes,
\begin{eqnarray}
\label{thermosimp}
F_t&=&\frac{2}{27}\left(9aS_1^2+2bS_1^3+3cS_1^4\right)
\end{eqnarray}

Which has stationary points when $dF_t/dS_1=0$ so that,
\begin{eqnarray}
S_1&=&0\\
S_1&=&\frac{1}{4c}\left(-b+\sqrt{b^2-24ac}\right)\label{Seqq}\\
S_1&=&\frac{1}{4c}\left(-b-\sqrt{b^2-24ac}\right)
\end{eqnarray}

This calculation can be performed using the $q_i$ variables if we first assume for simplicity that the laboratory frame of reference is set such that the directors are along the axes, as in Section~\ref{biaxial}. In this case we obtain the stationary points of the free energy as,
\begin{eqnarray}
q_1&=&0\\
q_1&=&\frac{1}{6c}\left(-b+\sqrt{b^2-24ac}\right)\\
q_1&=&\frac{1}{6c}\left(-b-\sqrt{b^2-24ac}\right)
\end{eqnarray}

By calculating $d^2F_t/dS_1^2$ and comparing the energies of each solution we find that
\begin{itemize}
\item $S_1=0$, the isotropic state, is globally stable for
$a>\frac{b^2}{27c}$,
  metastable for $0<a<\frac{b^2}{27c}$ and unstable for $a<0$.
\item $S_1=\frac{1}{4c}\left(-b+\sqrt{b^2-24ac}\right)$, the
nematic state, is globally stable
  for $a<\frac{b^2}{27c}$, metastable for
  $\frac{b^2}{27c}<a<\frac{b^2}{24c}$ and not defined for $a>\frac{b^2}{24c}$.
\item $S_1=\frac{1}{4c}\left(-b-\sqrt{b^2-24ac}\right)$ is
metastable (but has a negative value) for $a<0$, unstable for
$\frac{b^2}{27c}<a<\frac{b^2}{24c}$ and not
  defined for $a>\frac{b^2}{24c}$.
\end{itemize}
The equilibrium nematic scalar order parameter (\ref{Seqq}) will be denoted by $S_{eq}$.

There are clearly three important values of $a$: $a=\frac{b^2}{24c}$, the high temperature where the nematic state disappears; $a=\frac{b^2}{27c}$, the temperature at which the energy of the isotropic and nematic states are exactly equal; $a=0$, the low temperature where the isotropic state loses stability. If we use the notation $a=\alpha(T-T^*)$ then the critical temperatures are $T^+=\frac{b^2}{24\alpha c}+  T^*$; $T_{NI}=\frac{b^2}{27\alpha c}+T^*$; $T=T^*$.

\begin{figure}
\noindent \leavevmode \centerline{ 
\includegraphics[width=12cm]{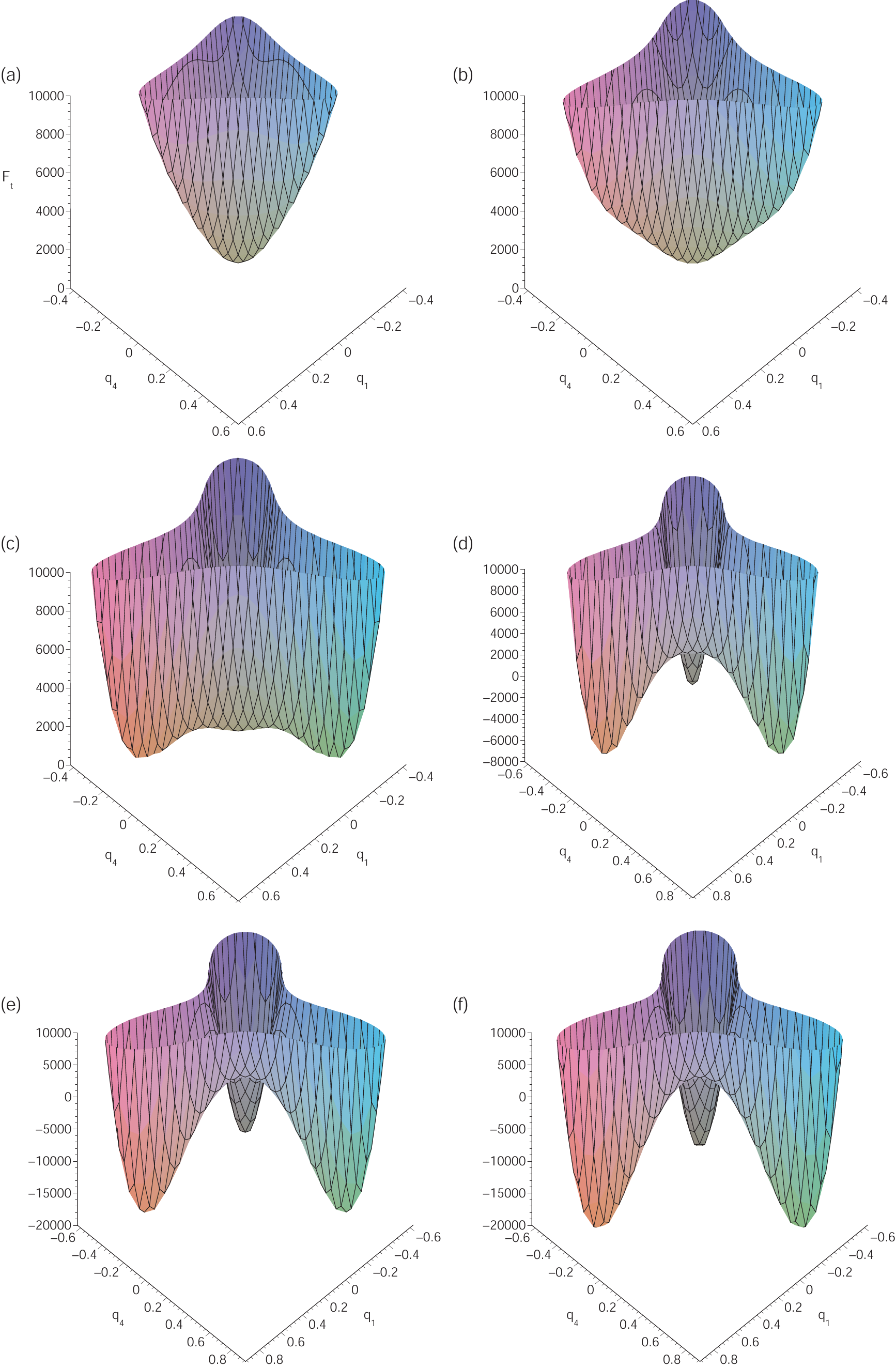}} \caption{The thermotropic potential for
parameters $\alpha=0.042\times 10^6$Nm$^{-2}$K$^{-1}$,
$b=-0.64\times 10^6$Nm$^{-2}$, $c=0.35\times 10^6$Nm$^{-2}$
\cite{Priestley} and (a) $\Delta T=2.0$, (b) $\Delta T=1.5$, (c)
$\Delta T=1.0$, (d) $\Delta T=0.5$, (e) $\Delta T=0.0$, (f)
$\Delta T=-0.1$. At high temperatures (a,b) only the isotropic
state is stable and at low temperature (d,e,f) the three uniaxial
states are stable. At intermediate temperatures (c) both the
isotropic and all three nematic states are stable.} \label{pot}
\end{figure}

\begin{figure}
\noindent \leavevmode \centerline{ 
\includegraphics[width=9cm]{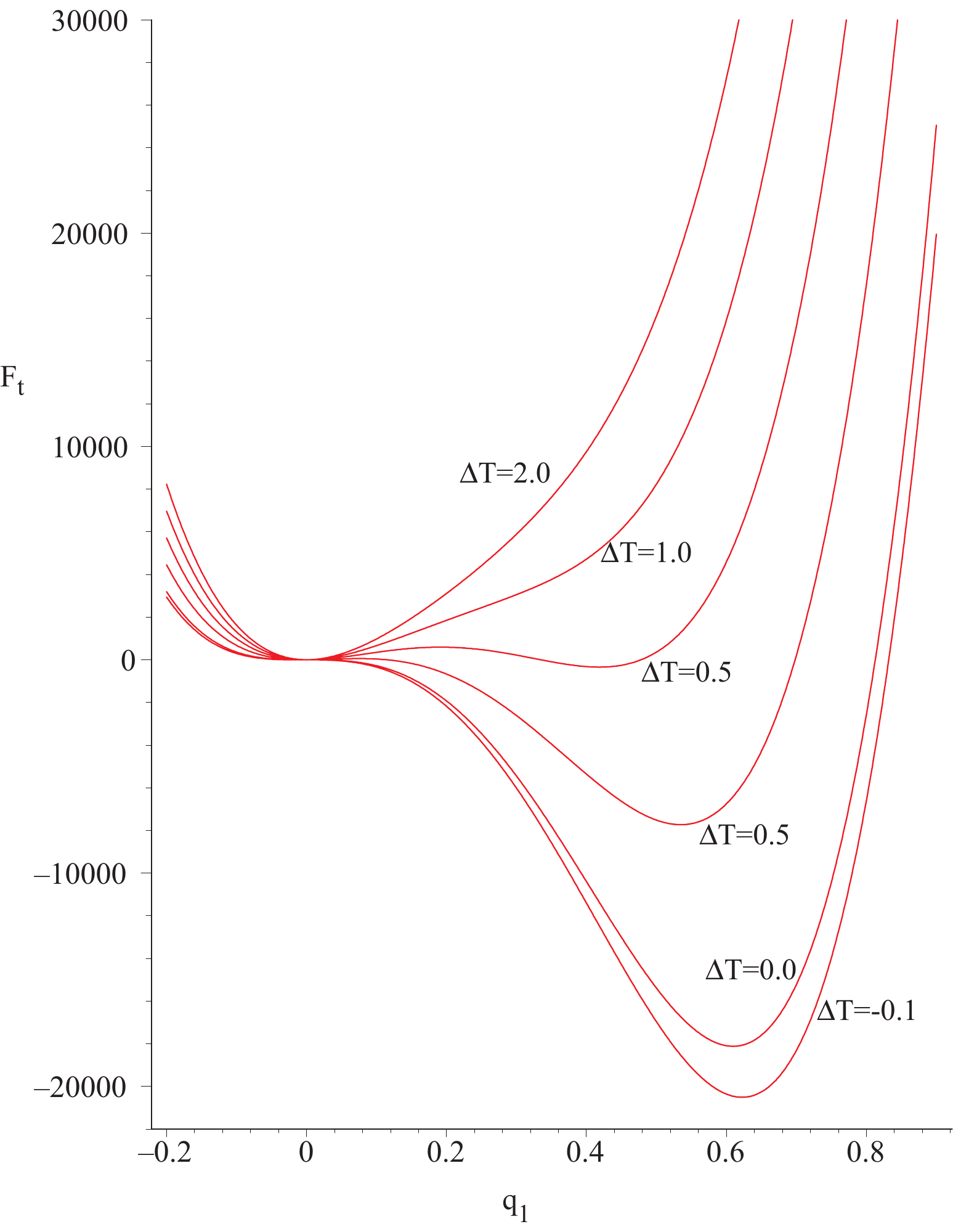}} \caption{Cross-section of Fig.~\ref{pot}
along the line $q_4=q_1/2$, the equivalent to the uniaxial state
$S_2=0$, for varying temperature with the parameter values
$\alpha=0.042\times 10^6$Nm$^{-2}$K$^{-1}$, $b=-0.64\times
10^6$Nm$^{-2}$, $c=0.35\times 10^6$Nm$^{-2}$ \cite{Priestley}. At
$\Delta T=0.5$ both the isotropic minima ($q_1=0$) and the
uniaxial nematic minima ($q_1\neq 0$) are locally stable.}
\label{potall}
\end{figure}

Therefore, for this thermotropic energy and depending on the
values of the coefficients $a$, $b$ and $c$, in an intermediate
temperature region the minimum at the isotropic state and the
minima at the uniaxial nematic states may all be locally stable.
However, at sufficiently low temperatures the isotropic state must
lose stability leaving only the uniaxial nematic states stable and
at sufficiently high temperatures the uniaxial states must lose
stability leaving only the isotropic state stable. The thermotropic potential function is shown in Figs~\ref{pot} and \ref{potall}. In these figures the tensor order parameter is assumed to be diagonalised so that the directors align with the laboratory frame of reference. Therefore, as mentioned in Section~\ref{Qtensor}, the off diagonal terms of ${\bf Q}$ are zero and $F_t$ depends only on $q_1$ and $q_4$. Figure~\ref{pot}(a) clearly shows the three uniaxial states and as temperature increases the potential function changes to exhibit a single minima at the isotropic state.

This expression for the thermotropic energy, eq.~(\ref{thermo}),
is essentially a Taylor series of the true thermotropic energy,
close to the point ${\bf Q}=0$. Therefore we must remember that
the Landau-de Gennes theory is only valid close to the
nematic-isotropic transition temperature, $T_{NI}$, where ${\bf
Q}\approx 0$. It is for this reason that we may assume that higher
order powers of ${\bf Q}$ may be neglected in eq.~(\ref{thermo}).
The first five powers of ${\bf Q}$ are included in
eq.~(\ref{thermo}) (the constant term is neglected as it will not
enter into a minimisation of the energy and the linear term is
taken to have zero coefficient since ${\rm tr}({\bf Q})=0$  so
that there is a minimum at ${\bf Q}=0$) since we assume there are
at most four minima in the potential function $F_t$.

\begin{figure}
\noindent \leavevmode \centerline{ 
\includegraphics[width=11cm]{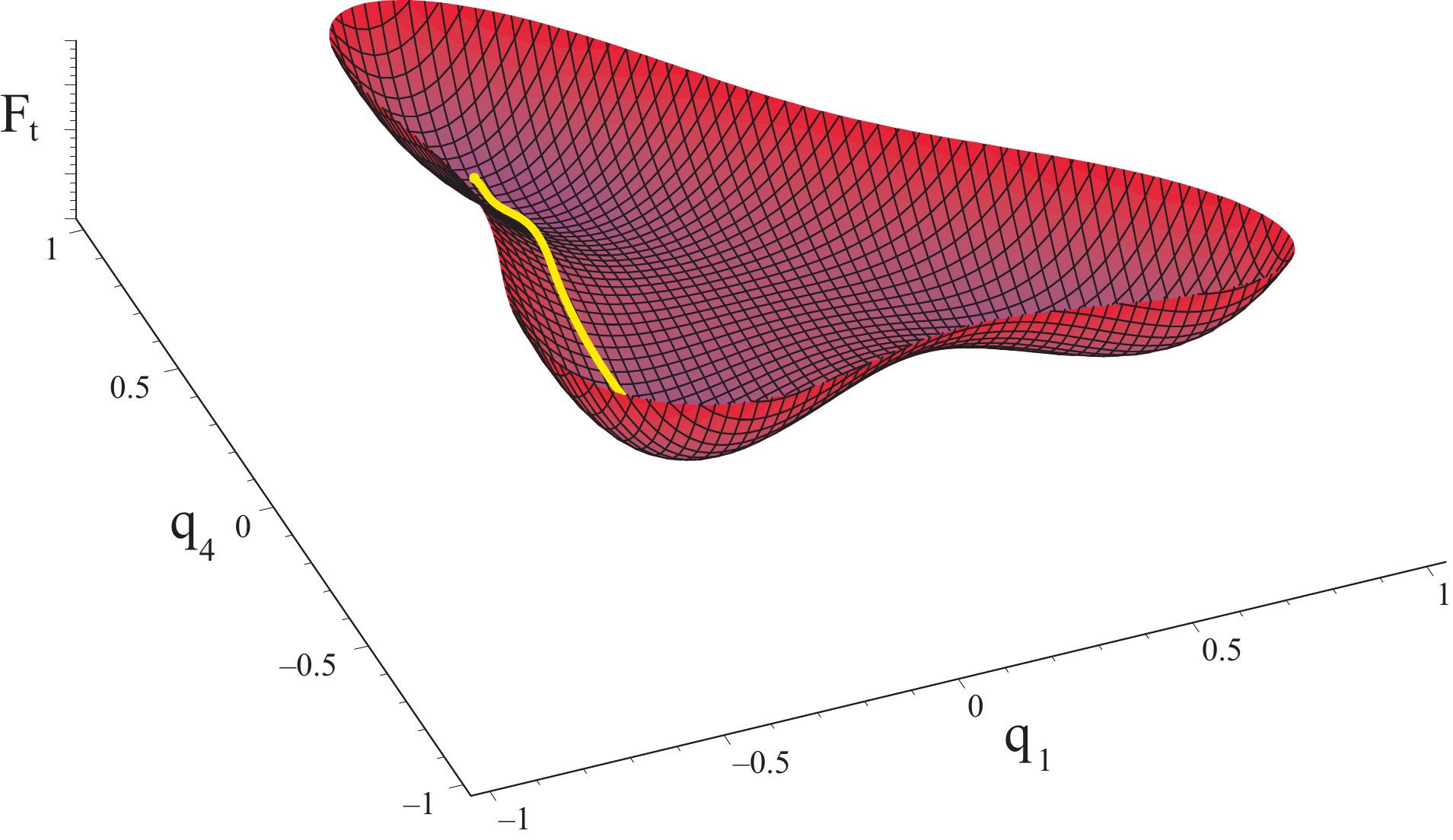}} \caption{A schematic plot of how a system
may be forced out of the thermotropic potential minimum. At one point in the
liquid crystal (represented by one end point of the yellow line)
the system is forced, through some external agent, to lie at one
of the uniaxial states. At another point in the liquid crystal
(represented by the other end point of the yellow line) the system
is forced to lie in a different uniaxial states. For the
macroscopic variables to be continuous through the region the
system continuously transforms from one state to the other, at
some points moving out of the potential minima. The yellow line
plots the state variables on the potential energy surface of the
system from one region in the liquid crystal to the other.}
\label{potdist}
\end{figure}

It would be energetically favourable for the system to lie in one
of the minima of $F_t$. When a liquid crystal material is forced
to contain some form of distortion (e.g. through the influence of surface or electric field
effects) there are two mechanisms for undertaking such a
distortion. For example imagine a region of liquid crystal
constrained to lie between two solid surfaces. If, through some
surface treatment, one of the surfaces forces the liquid crystal
in contact with that surface to lie in a fixed uniaxial state, i.e
in one of the minima of Fig.~\ref{pot}(a) whereas the other
surface forces the liquid crystal in contact with it to lie in a
different uniaxial state.
Firstly, for the system to move from one minima of $F_t$ to
another, the eigenvectors of ${\bf Q}$ may change. This is
equivalent to the molecular frame of reference changing. The
system effectively rotates the picture in Fig.~\ref{pot} so that
the system always remains in the same minima of $F_t$, it is
simply that the minima itself that changes position. Alternatively, the
eigenvalues of ${\bf Q}$ may change but the eigenvectors remain
fixed. This is equivalent to the picture in Fig.~\ref{pot}
remaining fixed and the system distorting from one uniaxial state
to another through a biaxial state, see Fig.~\ref{potdist}. Such a transformation is often termed eigenvalue exchange. Alternatively, there may be a combination of eigenvector and eigenvalue distortions and
exactly how the system distorts will depend on the competition
between the thermotropic energy $F_t$ and the elastic energy
$F_d$.

\subsection{Elastic energy}
\label{elastic} The distortional or elastic energy density, $F_d$,
of a liquid crystal is derived from the energy induced by
distorting the ${\bf Q}$ tensor in space. It is, generally,
energetically favourable for ${\bf Q}$ to be constant throughout
the material and any gradients in ${\bf Q}$ would lead to an
increase in distortional energy. $F_d$ therefore depends on the
spatial derivatives of {\bf Q}. Given a fixed distortion in space
of ${\bf Q}$ the distortional energy must remain unchanged if we
were to translate or rotate the material. Such restrictions (or
symmetries) mean that not all combinations of ${\bf Q}$
derivatives are allowed. In fact the elastic energy may be
simplified to \cite{Mori},
\begin{eqnarray}
\label{disten} F_d= \sum_{i=1,2,3\atop {j=1,2,3\atop
{k=1,2,3}}}&&\ \left[\frac{L_1}{2}\left(\frac{\partial
Q_{ij}}{\partial x_k}\right)^2+\frac{L_2}{2}\frac{\partial
Q_{ij}}{\partial x_j}\frac{\partial Q_{ik}}{\partial
x_k}+\frac{L_3}{2}\frac{\partial Q_{ik}}{\partial
x_j}\frac{\partial Q_{ij}}{\partial x_k}\right]\nonumber\\&&+
\sum_{i=1,2,3\atop {j=1,2,3\atop {k=1,2,3\atop
l=1,2,3}}}\left[\frac{L_4}{2}e_{lik}Q_{lj}\frac{\partial
Q_{ij}}{\partial x_k}+ \frac{L_6}{2}Q_{lk}\frac{\partial
Q_{ij}}{\partial x_l} \frac{\partial Q_{ij}}{\partial x_k}\right].
\end{eqnarray}

The coordinates $(x_1,\ x_2,\ x_3)=(x,\ y,\ z)$ are the usual
Cartesian coordinate system and $Q_{ij}$ is the $ij^{\rm th}$
element of ${\bf Q}$. The first four terms are quadratic in the
scalar order parameters $S_1$ and $S_2$ whereas the last term is
cubic in the scalar order parameters.

The elastic parameters $L_i$ are related to the Frank elastic
constants $k_{ij}$ by \cite{Mori}
\begin{eqnarray}
L_1&=&\frac{1}{6S_{exp}^2}(k_{33}-k_{11}+3k_{22}),\label{L1}\\
L_2&=&\frac{1}{S_{exp}^2}(k_{11}-k_{22}-k_{24}),\label{L2}\\
L_3&=&\frac{1}{S_{exp}^2}k_{24},\label{L3}\\
L_4&=&\frac{2}{S_{exp}^2}q_0k_{22},\label{L4}\\
L_6&=&\frac{1}{2S_{exp}^3}(k_{33}-k_{11}),\label{L6}
\end{eqnarray}
where $S_{exp}$ is the uniaxial order parameter of the liquid
crystal when the experimental measurement of the $L_i$ was taken
and may not be equal to the current order parameter
$S_{eq}=\left(-b+\sqrt{b^2-24ac}\right)/(4c)$ mentioned in
Section~\ref{LdeG}. The parameter $q_0$ is the chirality of the
liquid crystal and if an achiral liquid crystal is under
consideration then $L_4=0$.

It has been shown that there are seven elastic terms of cubic
order \cite{Berreman, Schiele} but we will only include one, the
$L_6$ term in eq.~(\ref{disten}), in order to ensure we can model
a nematic state with non-equal elastic constants $k_{11}$,
$k_{22}$, $k_{33}$. Without the $L_6$ term we have four elastic
parameters $L_1,\ldots,L_4$ in the ${\bf Q}$ tensor elastic energy
but we have five independent parameters in the Frank approach,
$k_{11}$, $k_{22}$, $k_{33}$, $k_{24}$ and $q_0$. Including the
$L_6$ term removes the degeneracy in the mapping from the ${\bf
Q}$ tensor to the Frank energy approaches.

\subsection{Electrostatic energy}
\label{electro} The liquid crystal will interact with an
externally applied electric field or indeed self-induce an
internal electric field due to the dielectric and spontaneous
polarisation effects. The electrostatic energy density is
\begin{equation}
F_e=-\int {\bf D}.{\rm d}{\bf E},
\end{equation}
where ${\bf D}$ is the displacement field and ${\bf E}$ is the
electric field. The constitutive equation,
\begin{eqnarray}
{\bf D}&=&\epsilon_0{\bf E}+{\bf P}\nonumber\\
&=&\epsilon_0{\bf E}+{\bf P}_i+{\bf P}_s\nonumber\\
&=&\epsilon_0{\bf E}+\epsilon_0{\pmb \chi}{\bf E}+{\bf P}_s\nonumber\\
&=&\epsilon_0{\pmb \epsilon}{\bf E}+{\bf P}_s,
\end{eqnarray}
relates the displacement field with the electric field ${\bf E}$
and the internal polarisation ${\bf P}$. This polarisation is
split into dielectrically induced polarisation ${\bf P}_i$ and the
spontaneous polarisation ${\bf P}_s$. The induced polarisation is
dependent on the dielectric susceptibility $\chi$ and the electric
field ${\bf E}$ and the dielectric tensor is then defined as
$\epsilon={\bf I}+\chi$ where ${\bf I}$ is the identity matrix.
The electrostatic energy density is then,
\begin{equation}
\label{electrostatic} F_e=-\int (\epsilon_0{\pmb
  \epsilon}{\bf E}+{\bf P}_s).{\rm d}{\bf
  E}=-\frac{1}{2}\epsilon_0\left({\pmb
  \epsilon}{\bf E}\right).{\bf E}-{\bf P}_s.{\bf E}.
\end{equation}

In a nematic liquid crystal the dielectric tensor is approximated
to ${\pmb \epsilon}=\Delta\epsilon^*{\bf Q}+\bar{\epsilon}{\bf I}$
where $\Delta\epsilon^*=(\epsilon_{||}-\epsilon_\perp)/S_{exp}$ is
the scaled dielectric anisotropy and
$\bar{\epsilon}=(\epsilon_{||}+2\epsilon_\perp)/3$ is an average
permittivity. Within an isotropic material, such as an alignment
layer of photo-resist material, the dielectric tensor is diagonal
and isotropic, $\epsilon=\epsilon_I {\bf I}$.

In the present, nematic, situation a spontaneous polarisation
vector is assumed to derive only from a flexoelectric type of
polarisation, that is a polarisation caused by a distortion of the
molecular arrangement. This may be due to a shape asymmetry in the
molecules or due to a distortion of a pair-wise coupling of
molecules. For either of these mechanisms, the polarisation may be
written, to leading order, in terms of the ${\bf Q}$ tensor as
\cite{Alexe},
\begin{equation}
\label{flexo} {\bf P}_s=\bar{e}\nabla. {\bf Q},
\end{equation}
where the $i^{\rm th}$ component of $\nabla. {\bf Q}$ is
understood to be $\sum_{j=1,2,3}{\partial Q_{ij}}/{\partial x_j}$.

If we consider the flexoelectric polarisation of a uniaxial state,
where ${\bf Q}=S\left(\left({\bf n}\otimes{\bf n}\right)-{\bf
I}/3\right)$, we find
\begin{equation}
{\bf P}_s=\bar{e}\left(S\left(\nabla . {\bf n}\right)\,{\bf
n}+S\left(\nabla\times{\bf n}\right)\times{\bf n}+\left({\bf
n}.\nabla S\right)\,{\bf n}-\frac{1}{3}\nabla S\right).
\end{equation}
Comparing this to the standard expressions for flexoelectric
polarisation and order electricity \cite{Meyer,deGennes,Barbero},
\begin{eqnarray}
{\bf P}_f&=&e_{11}\left(\nabla . {\bf n}\right)\,{\bf
n}+e_{33}\left(\nabla\times{\bf n}\right)\times{\bf n},\label{f1} \\
{\bf P}_o&=&r_{1}\left({\bf n}.\nabla S\right)\,{\bf
n}+r_{2}\nabla S,\label{o1}
\end{eqnarray}
we see that eq.~(\ref{flexo}) is equivalent to eqs~(\ref{f1}) and
(\ref{o1}) when we set $e_{11}=e_{33}=S\bar{e}$ and $r_1=\bar{e}$,
$r_2=-\bar{e}/3$. Therefore, without taking higher order terms in
eq.~(\ref{flexo}), this expression assumes that the coefficients
of flexoelectric and order electric terms are equal. Within a
region where $S$ is constant, only the flexoelectric polarisation
eq.~(\ref{f1}) would be present.

In order to distinguish between the flexoelectric parameters
$e_{11}$ and $e_{33}$ higher order terms are needed in the
flexoelectric polarisation term in eq.~(\ref{flexo})
\cite{Osipov1}. For instance, if we include second order terms the
$i^{\rm th}$ component of the polarisation vector is
\begin{eqnarray}
({\bf P}_{s})_{i}=p_1 \sum_{j=1,2,3}\frac{\partial
Q_{ij}}{\partial x_j}+p_2 \sum_{j=1,2,3\atop
{k=1,2,3}}Q_{ij}\frac{\partial Q_{jk}}{\partial x_k} +p_3
\frac{\partial }{\partial x_i}\left(\sum_{j=1,2,3\atop
{k=1,2,3}}Q_{jk}Q_{jk}\right)+p_4 \sum_{j=1,2,3\atop
{k=1,2,3}}Q_{jk}\frac{\partial Q_{ji}}{\partial x_k}.
\end{eqnarray}
For a uniaxial material we can substitute, expand and collect
terms as before. This gives
\begin{eqnarray}
e_{11} & = & S p_1 +\frac{S^2}{3}(2p_2-p_4), \nonumber \\
e_{33} & = & S p_1 +\frac{S^2}{3}(2p_4-p_2), \nonumber \\
r_1 & = & p_1+\frac{S}{3}(p_2+p_4), \nonumber \\
r_2 & = & \frac{1}{3}\left(-p_1+\frac{S}{3}(p_2+p_4)+4S
p_3\right). \nonumber
\end{eqnarray}
Therefore, including second order terms does allow us to have
different values for $e_{11}$ and $e_{33}$ if $p_2\ne p_4$. However,
deciding on values for $p_1\dots p_4$ is non-trivial. Barbero et
al.~\cite{Barbero} have stated that $p_1$, $p_2$ and $p_4$ can be
obtained by measuring the flexoelectric parameters as a function
of temperature, and assume a similar magnitude for $p_3$, but
there seems to be no such measurements in the literature.

At this point the electric field within the whole cell is unknown
but may be found using Maxwell's equations which are, in this
static case with no free charges,
\begin{eqnarray}
\nabla.{\bf D}&=&0,\label{max1}\\
\nabla\times {\bf E}&=&0.\label{max2}
\end{eqnarray}
Equation~(\ref{max2}) means that we may define a scalar function
$U$, the electric potential, which satisfies ${\bf E}=-\nabla\,U$ (it is convention to make ${\bf E}$ the negative gradient of $U$).
If we find the function $U$  then eq.~(\ref{max2}) is
automatically satisfied. Equation~(\ref{max1}) is then the
governing equation for the electric potential  $U$,
\begin{eqnarray}
0&=&\nabla.{\bf D}=\nabla.\left(-\epsilon_0{\pmb
\epsilon}\nabla\,U+{\bf P}_s\right),\label{max3}
\end{eqnarray}
and the free energy density,
\begin{equation}
\label{feelec} F_e=-\frac{1}{2}\epsilon_0{\pmb
\epsilon}\nabla\,U.\nabla\,U+{\bf P}_s.\nabla\,U,
\end{equation}
will enter the total free energy density to be minimised in order
to obtain the Euler-Lagrange equations for the $q_i$.
Equation~(\ref{max3}) is in fact equivalent to the Euler-Lagrange
equation derived from minimising the free energy in
eq.~(\ref{feelec}) with respect to $U$.

At any material boundary (i.e. between the liquid crystal and a substrate) we must ensure that the standard conditions of electrostatics are obeyed, that the electric potential $U$ continuous, the component of the displacement field
normal to the boundary is continuous and the component of the
electric field parallel to the boundary is continuous. The external boundary conditions for $U$ are usually set at the electrodes
where, for example, one electrode is set to be earthed and so
$U=0$ and the other electrode is set to be a fixed voltage $U=V$.

\subsection{Surface energy}
\label{surface1} When the liquid crystal molecules are close to a
solid surface they will feel a molecular interaction force.
Whatever this force is we would like to model this liquid
crystal-surface interaction in a macroscopic framework. That is,
how does the surface interact with the macroscopic variables (the
elements of the tensor order parameter ${\bf Q}$). If the surface
is treated is some way, usually by coating the surface with a
chemical and possibly rubbing the surface in a fixed direction so
as to create a preferred surface direction, then we will assume
that at that surface the directors ${\bf n}$ and ${\bf m}$ would
prefer to lie in a certain direction. We may also assume that the
solid surface may affect the amount of order (i.e.~the variables
$S_1$ and $S_2$) at that surface. Associated with this preference
for a certain orientation and order will be a {\it surface
energy}, ${\cal F}_{surface}$, which will have a minimum at the
preferred state. This surface energy will be a function of the
value of the dependent variables at the surface. Thus, if $\cal S$
is the surface in contact with the liquid crystal, ${\cal
F}_{surface}={\cal
  F}_{surface}({\bf Q}|_{\cal S})$. If the variables
are forced to move out of the minima, usually by the bulk of the
liquid crystal being in some alternate state, then the surface
energy will increase and there will be a competition between surface
energy and bulk energy. In equilibrium, a balance will be reached
such that the total energy in eq.~(\ref{totenergyQ}) is minimised.

{\bf Weak anchoring}:
One simple form of the free energy density, $F_{s}$, has a single
minimum at the point where the dependent variables take the value
dictated by the surface treatment,
\begin{equation}
\label{align}
 F_{s}=\frac{W}{2}{\rm tr}\left({\bf Q}|_{\cal S}-{\bf Q}_s\right)^2
\end{equation}
where ${\bf Q}_s$ is the value of the tensor order parameter
preferred by the surface and $W$ is the anchoring energy. When the
system is in equilibrium the calculus of variations gives the set
of boundary conditions in eqs~(\ref{qeqsurf2}) which are
similar to a torque balance at the surface, of the distortion
torque and the torque due to the surface energy function. If we
were to compare this energy to a Rapini-Papoular type anchoring
energy where, say, a preference for the director to lie in the $x$
direction is given by the energy density
$F_{s}=\frac{W_s}{2}\sin^2\theta$, then the relationship between
the Rapini-Papoular anchoring strength $W_s$ and the ${\bf Q}$
tensor anchoring strength in eq.~(\ref{align}) will be,
\begin{equation}
W=\frac{W_s}{2{S_s}^2},
\end{equation}
where $S_s$ is the preferred surface order parameter.

An example of the type of anchoring in eq.~(\ref{align}) is if we
have a surface which prefers the nematic to be uniaxial with the
director in the $z$ direction and scalar order parameter of $0.6$.
We then take $S_2=0$, $\theta=\pi/2$ and $S_1=0.6$ in
eqs~(\ref{qs1}-\ref{qs5}) so that the preferred ${\bf Q}$ tensor
is,
\begin{equation}
\label{qmatsurf} {\bf Q}_s=\left(
\begin{array}{ccc}
-0.2&0&0\\ 0&-0.2&0\\ 0&0&0.4
\end{array}
\right),
\end{equation}
or in terms of the $q_i$ values,
$q_1=-0.2,\,q_2=0.0,\,q_3=0.0,\,q_4=-0.2,\,q_5=0.0$.

Another important example of weak anchoring is homeotropic
anchoring on a non-planar surface. Such anchoring prefers the
director to lie perpendicular to the surface at all points. If we
denote the unit normal to the surface as
${\pmb{\nu}}=(\nu_x,\nu_y,\nu_z)$ then this will be the preferred
director. If we assume that the surface will prefer a uniaxial
state (as suggested by the local symmetry of homeotropic
anchoring) then we may use eq.~(\ref{q}) with ${\bf n}={\pmb
\nu}$, $S_1=S_s$ and $S_2=0$ to obtain the preferred ${\bf Q}$
tensor,
\begin{equation}
\label{qmatsurf2} {\bf Q}_s=S_s\left(
\begin{array}{ccc}
{\nu_x}^2-\frac{1}{3}
&\nu_x\nu_y&\nu_x\nu_z\\ \nu_x\nu_y&{\nu_y}^2-\frac{1}{3}&\nu_y\nu_z\\
\nu_x\nu_z&\nu_y\nu_z&\frac{2}{3}-{\nu_x}^2-{\nu_y}^2
\end{array}
\right).
\end{equation}

{\bf Strong/infinite anchoring}: The case of strong anchoring mentioned in Section \ref{theory} is
equivalent to the limit $W_s\rightarrow \infty$, and is therefore often termed infinite anchoring. In this case the
Dirichlet condition
\begin{equation}
{\bf Q}={\bf Q}_s,
\end{equation}
is applied on the boundary instead of the weak anchoring condition
for which a surface energy is minimised.

{\bf Planar degenerate anchoring}: 
\label{surface2} Another common liquid crystal substrate (i.e.~untreated SU8)
leads to planar degenerate anchoring. In this situation the preferred
orientation for the directors is to lie parallel to the substrate.
There is no preference as to which direction on the plane of the
surface to lie but simply that the director lies on the surface.

The most general surface energy density in this case is
\cite{Osipov},
\begin{eqnarray}
\label{planarbc} F_{s}&=&c_1{\pmb \nu}.{\bf Q}.{\pmb \nu}+
c_2({\pmb \nu}.{\bf Q}.{\pmb \nu})^2+c_3{\pmb \nu}.{\bf Q}^2.{\pmb
  \nu}\nonumber\\
&& \hspace{1cm}+a_s\,{\rm tr}\left({\bf
Q}^2\right)+\frac{2b_s}{3}\,{\rm tr}\left({\bf
Q}^3\right)+\frac{c_s}{2}\,\left({\rm tr}\left({\bf
Q}^2\right)\right)^2.
\end{eqnarray}

In this energy density the first three terms give the most general
energy density, up to quadratic order, which specifies that the
eigenvectors of ${\bf Q}$ lie parallel to the surface with normal
${\pmb \nu}$. The last three terms in eq.~(\ref{planarbc}) are
added to specify preferred eigenvalues of ${\bf Q}$.

If we assume the liquid crystal has taken a uniaxial state at the
surface, $S_1=S$ and $S_2=0$, then the planar degenerate surface
energy density is
\begin{equation}
F_s=\frac{S}{3}\left(3c_2S\sin^4\theta+\left(3c_1+(c_3-2c_2)S\right)\sin^2\theta\right)+f(S),
\end{equation}
which has a minimum at $\theta=0$ when $S(3c_1+(c_3-2c_2)S)>0$.
The effective anchoring strength, when compared to a
Rapini-Papoular energy, may be thought of as
$W_s=\frac{2}{3}S(3c_1+(c_3-2c_2)S)$. It is clear that there is no
dependence on the azimuthal angle $\phi$ as such a degenerate
anchoring condition would suggest.

With all terms in the free energy now specified, as well as appropriate examples of boundary conditions, it is possible to apply the governing equations to a realistic problem.

\section{Example: The Zenithal Bistable Device}
\label{example}
\begin{figure}[th]
\noindent \leavevmode \centerline{ 
\includegraphics[width=7cm]{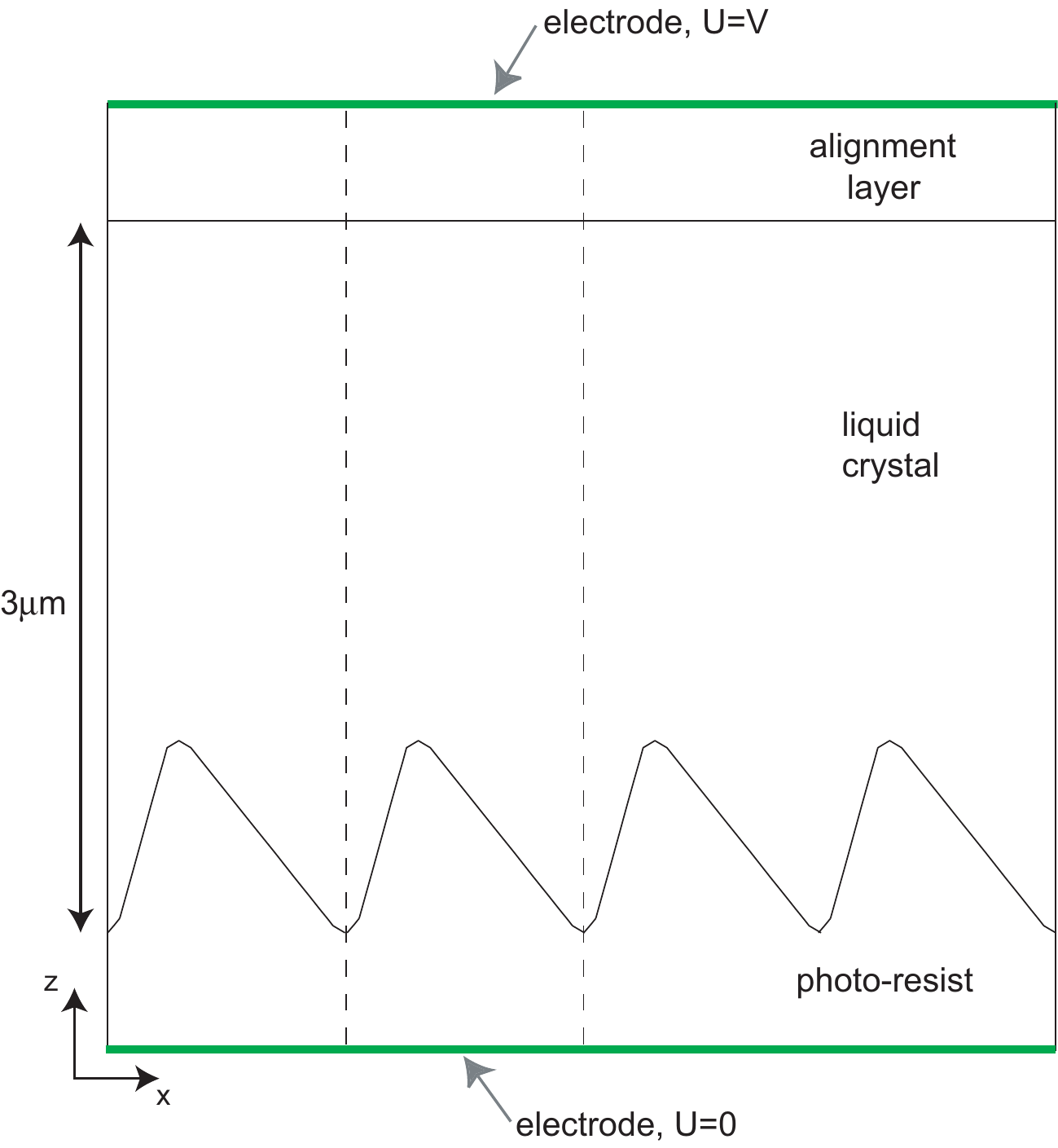}} \caption{Typical ZBD cell. A periodic grating
structure formed out of a photo-resist material. The upper surface
is coated with an alignment layer material and electrodes are
placed around the system. In the numerical computation we assume
that we may consider only one of the repeating cells and enforce
periodic boundary conditions.} \label{ZBDfig}
\end{figure}

Consider a typical ZBD system, such as in Fig.~\ref{ZBDfig}, as an
example \cite{Brown,Newton}. A layer of liquid crystal is
sandwiched between an alignment layer on the upper surface and a
patterned photo-resist surface on the lower substrate, around
which are two electrodes. The Euler-Lagrange equations from
eq.~(\ref{qeq}), with energy densities
eqs~(\ref{thermo}),~(\ref{disten}),~(\ref{electrostatic}), for the
five ${\bf Q}$ tensor elements will be solved in the liquid
crystal region with boundary conditions at the alignment layer
and the photo-resist interfaces. At both the alignment layer
surface and the photoresist surface we will assume that there
exists homeotropic anchoring so that the preferred ${\bf Q}$
tensor is a uniaxial tensor with a director in the surface normal
direction. We therefore use the boundary condition in
eq.~(\ref{qeqsurf2}) and the surface energy in eq.~(\ref{align})
with eq.~(\ref{qmatsurf2}).

The electric potential $U$ will be solved within the whole region
bounded by the two electrodes. Within the whole region $U$ is
subject to eq.~(\ref{max3}) where the dielectric tensor is
different within the alignment layer, the liquid crystal layer and
the photo-resist. Within the alignment layer
$\epsilon=\epsilon_{a}{\bf I}$, in the liquid crystal layer
$\epsilon=\Delta\epsilon^*{\bf Q}+\bar{\epsilon}{\bf I}$ and
within the photo-resist $\epsilon=\epsilon_{p}{\bf I}$. The
boundary conditions for $U$ are $U=0$ at the bottom electrode and
$U=V$ at the top electrodes.

We will also apply periodic boundary conditions on two $yz$ planes
(dashed lines in Fig.~\ref{ZBDfig}) in order to model an infinite
periodic grating. In this example no voltage is applied to the
electrodes.

\begin{figure}
\noindent \leavevmode \centerline{ 
\includegraphics[width=14cm]{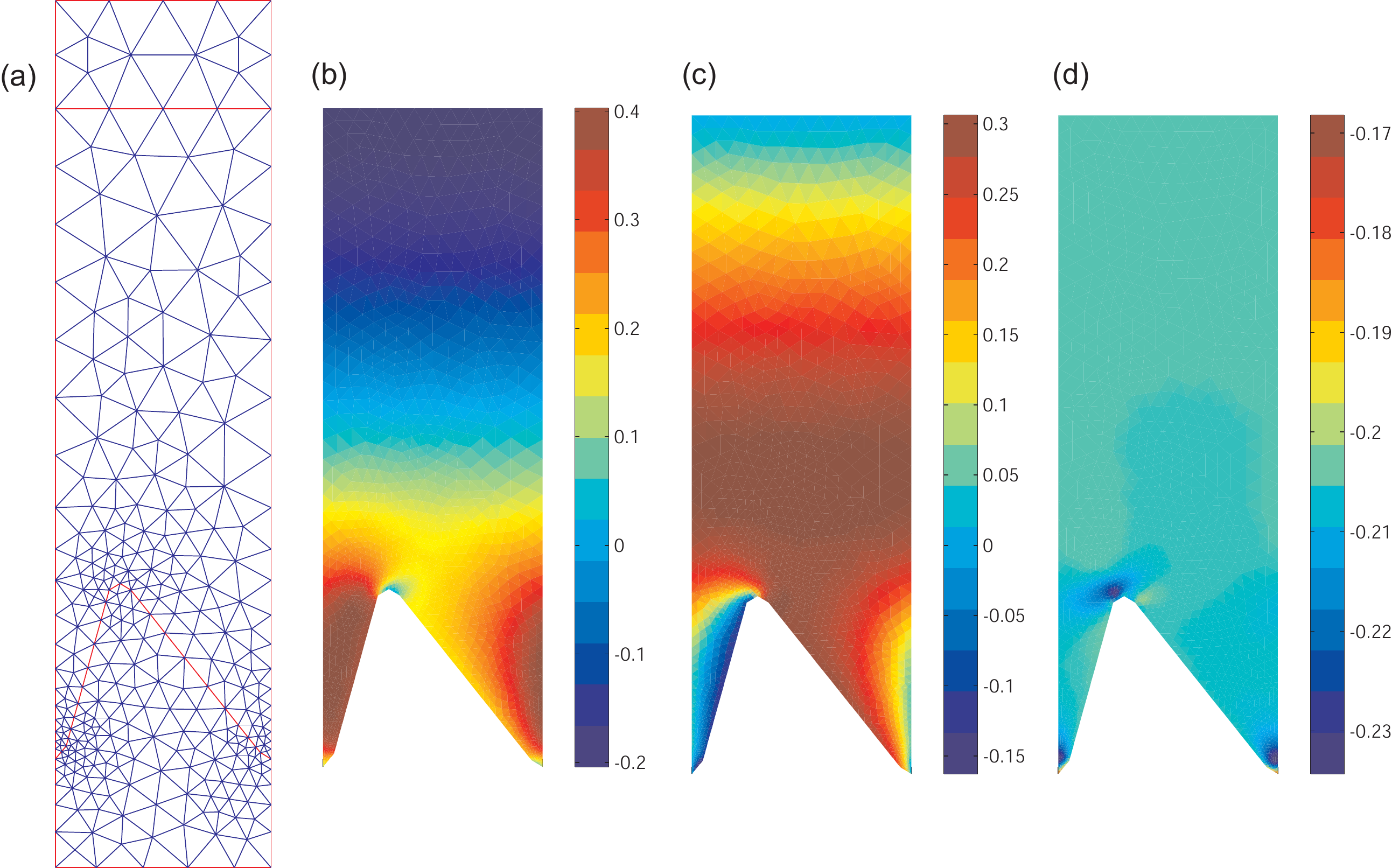}} \caption{ZBD display with zero voltage, the HAN
state: (a) finite element mesh and the solutions for (b) $q_1$,
(c) $q_2$, (d) $q_4$.} \label{qfig}
\end{figure}

\begin{figure}
\noindent \leavevmode \centerline{ 
\includegraphics[width=14cm]{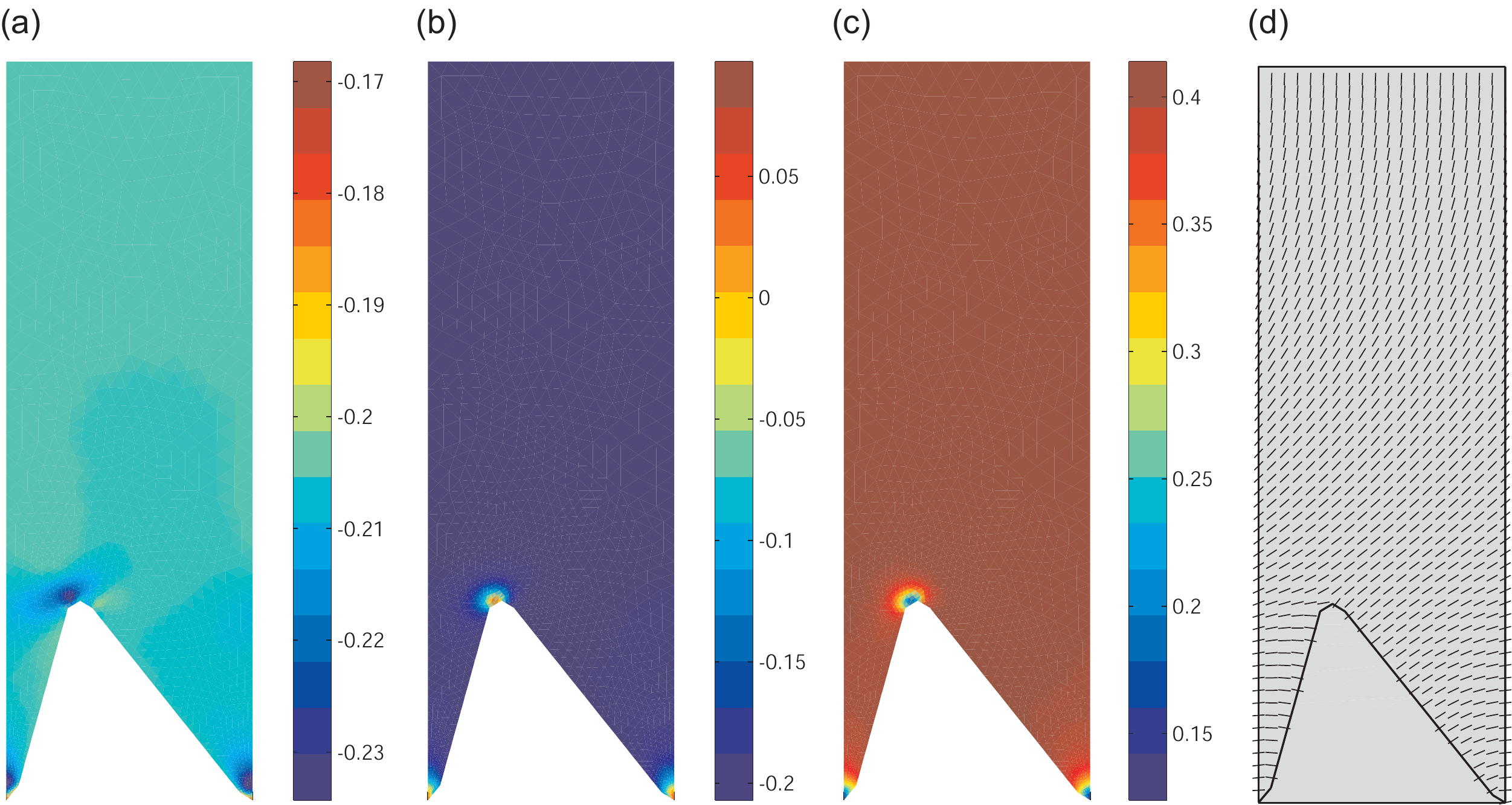}} \caption{ZBD display with zero voltage, the HAN
state: the three eigenvalues (a), (b) and (c) and the eigenvector
associated with the largest eigenvalue.} \label{eigfig}
\end{figure}

Using MATLAB\cite{MATLAB:2013} and the finite element modelling package COMSOL\cite{COMSOL} the
domain is meshed (see Fig.~\ref{qfig}(a)) and the solution of the
governing equations is found. As an example Figure~\ref{qfig}
shows the solutions for $q_1$, $q_3$ and $q_4$. The variables
$q_2$, $q_5$ and $U$ are identically zero through the region.
Possibly more instructive are plots of the eigenvalues and
eigenvectors of ${\bf Q}$, see Figure~\ref{eigfig}.

From Fig.~\ref{eigfig} we see that there exist two regions where
the eigenvalues vary significantly. Near the peak and trough of
the grating structure there exist -1/2 and +1/2 defects
respectively. This defect structure means that in the region
immediately above the grating the director (the eigenvector in
Fig.~\ref{eigfig}(d)) is almost horizontal and a hybrid aligned
nematic (HAN) type structure results.

\section{Summary}
In this paper we have introduced a simple form of the ${\bf Q}$-tensor theory for calamitic nematic liquid crystals. We have described the governing equations for a model where fluid flow is neglected and given the forms for the constitutive components of energy and dissipation. This paper should hopefully allow both simple numerical models of liquid crystal devices to be constructed and give the reader a better understanding of the vast array of publications written in the last 10-15 years which have utilised such a theory. 

One final note of caution is necessary. A theory which allows for the presence of defects is always going to be relatively difficult to solve numerically. The large difference in spatial scale between a device geometry (on the scale of 10s or 100s of microns) and the defect core (on the scale of 10s to 100s of nanometres) means that discretisation of even a static example can lead to large numbers of numerical elements. A non-uniform grid is essential and, if the dynamic equations are to be solved, both adative meshing and timestepping will also be necessary. A number of papers have been written about these topics and those attempting to model such systems would be advised to consult these papers or seek assistance from researchers in the area\cite{Ramage, Macdonald, Macdonald2}.


\begin{thebibliography}{99}
\bibitem{Virga} E. G. Virga, {\it Variational theories for liquid
crystals}, Chapman \& Hall, London (1994).
\bibitem{SonnetVirga} A. M. Sonnet and E. G. Virga, {\it Dissipative Ordered Fluids: Theories for Liquid Crystals}, Springerl, New York (2010).
\bibitem{biax1} L. A. Madsen, T. J. Dingemans,
M. Nakata, and E. T. Samulski, {\it Thermotropic biaxial nematic
liquid crystals},  Phys. Rev. Lett. {\bf 92}, 145505 (2004).
\bibitem{biax2} B. R. Acharya, A. Primak, and S. Kumar, {\it Biaxial nematic
phase in bent-core thermotropic mesogens}, Phys. Rev. Lett. {\bf
92}, 145506 (2004).
\bibitem{Luckhurst:2012} G. R. Luckhurst, S. Naemura, T. J. Sluckin, S. K. Thomas, \& S. S. Turzi, {\it Molecular-field-theory approach to the {L}andau
  theory of liquid crystals: {U}niaxial and biaxial nematics}, Phys.\ Rev.\ E {\bf 85}, {031705} (2012).
\bibitem{Schophol} N. Schophol and T. J. Sluckin, {\it Defect core structure in nematic liquid
crystals}, Physical Review Letters {\bf 59}, p.2582 (1987).
\bibitem{Priestley} E. B. Priestley, P. J. Wojtowicz and P. Sheng, {\it Introduction to liquid
crystals}, Plenum Press, New York and London (1976).
\bibitem{Mori} H. Mori, E. C. Gartland, J. R. Kelly and P. J. Bos,
{\it Multidimensional director modeling using the Q tensor
representation in a liquid crystal cell and its application to the
$\pi$ cell with patterned electrodes}, Jap. J. App. Phys. {\bf
38}, p.135 (1999).
\bibitem{Berreman} D. W. Berreman and S. Meiboom, {\it Tensor
representation of Oseen-Frank strain energy in uniaxial
cholesterics}, Phys. Rev. A {\bf 30}, p.1955 (1984).
\bibitem{Schiele} K. Schiele and S. Trimper,
{\it On the elastic constants of a nematic liquid crystal},
Physica Status Solidi B {\bf 118}, p.267 (1983).
\bibitem{Alexe} A. L. Alexe-Ionescu, {\it Flexoelectric
polarisation and 2nd order elasticity for nematic liquid
crystals}, Phys. Lett. A {\bf 180}(6), p.456 (1993).
\bibitem{Meyer} R.
Meyer, {\it Piezoelectric effects in liquid crystals}, Phys. Rev.
Lett. {\bf 22}, p.918 (1969).
\bibitem{deGennes}
P. G. de Gennes and J. Prost, {\it The physics of liquid
crystals}, 2nd Edition, Clarendon Press, Oxford (1993).
\bibitem{Barbero} G. Barbero, I. Dozov, J. F. Palierne and G.
Durand, Phys. Rev. Lett. {\bf 56}, p.2056 (1986).
\bibitem{Osipov1} M. A. Osipov, {\it The order parameter dependence
of the flexoelectric coefficients in nematic liquid-crystals}, J.
Phys. Lett. (Paris) {\bf 45}, p.L823 (1984).
\bibitem{Osipov} M. A. Osipov and S. Hess, {\it Density functional
approach to the theory of interfacial properties of nematic liquid
crystals}, Journal of Chemical Physics {\bf 99}, p.4181 (1993).
\bibitem{Brown} C. V. Brown, G. P. Bryan-Brown, and J. C. Jones, U.S. Patent No.
6,249,332 19 June 2001.
\bibitem{Newton} C. J. P. Newton and T. P.
Spiller, {\it Bistable Nematic Liquid-Crystal Device Modeling}, in
SID Proceedings of IDRC 97, edited by J. Morreale SID, Santa Ana,
CA, p.13 (1997).
\bibitem{MATLAB:2013} MATLAB version 8.1 (R2013a), The MathWorks Inc., Natick, Massachusetts
(2013)
\bibitem{COMSOL} COMSOL Multiphysics, version 4.3b, COMSOL Inc. (2013)
\bibitem{Ramage} A. Ramage and E. C. Gartland Jnr, {\it A preconditioned nullspace method for liquid crystal director modelling}, SIAM Journal on Scientific Computing {\bf 35}, p. B226 (2013).
\bibitem{Macdonald} C. S. MacDonald, J. A. Mackenzie, A. Ramage and C. J. P. Newton, {\it Efficient moving mesh methods for Q-tensor models of nematic liquid crystals}, Strathclyde Mathematics Research Report No. 10 (2013).
\bibitem{Macdonald2} MacDonald, C. S. MacDonald, J. A. Mackenzie, A. Ramage and C. J. P. Newton, {\it Robust adaptive computation of a one-dimensional Q-tensor model of nematic liquid crystals}, Computers and Mathematics with Applications, {\bf 64}(11), p. 3627 (2012).
\end{thebibliography}
\end{document}